\documentclass{article}

\usepackage[preprint]{neurips_2025}

\usepackage[utf8]{inputenc} % allow utf-8 input
\usepackage[T1]{fontenc}    % use 8-bit T1 fonts
\usepackage{hyperref}       % hyperlinks
\usepackage{url}            % simple URL typesetting
\usepackage{booktabs}       % professional-quality tables
\usepackage{amsfonts}       % blackboard math symbols
\usepackage{nicefrac}       % compact symbols for 1/2, etc.
\usepackage{microtype}      % microtypography
\usepackage{xcolor}         % colors
\usepackage{multirow}
\usepackage{graphicx}
\usepackage{todonotes}
\usepackage{pifont}
\usepackage{tcolorbox}
\usepackage[normalem]{ulem}
\usepackage{amsmath}
\usepackage{amssymb}
\usepackage{mathtools}
\usepackage{amsthm}
\usepackage{placeins}
\usepackage{inconsolata}
\usepackage{multirow}
\usepackage{enumitem}

\usepackage{enumitem}

\title{How does longer temporal context enhance multimodal narrative video processing in the brain?}

\author{}

\author{
Subba Reddy Oota$^{1}$\thanks{Equal contribution}, Anant Khandelwal$^{2}$\footnotemark[1], Prachi Jindal$^{3}$  \\ \textbf{Manish Gupta}$^4$, \textbf{Bapi S. Raju}$^5$,   \textbf{Tanmoy Chakraborty}$^3$ \\
\normalsize $^1$Technische Universität Berlin, Germany, $^2$Microsoft Research, Bangalore, India \\
$^3$IIT Delhi, India, $^4$Microsoft, India, $^5$IIIT-Hyderabad, India
 \\
 \texttt{\small subba.reddy.oota@tu-berlin.de, \{anantk, gmanish\}@microsoft.com} \\
\small \texttt{prachi.jindal9561@gmail.com, raju.bapi@iiit.ac.in,  tanchak@iitd.ac.in}
}

\begin{document}

\maketitle

\begin{abstract}
% Understanding how humans and artificial intelligence systems process complex narrative videos is a fundamental challenge at the intersection of neuroscience and machine learning. This study investigates how the temporal context length of video clips (from brief 3\,s segments to longer 12\,s excerpts) influences semantic and contextual information extraction by state-of-the-art multimodal large language models (MLLMs) trained on video and audio data. Using functional MRI recordings of participants watching full-length movies, we examine how brain regions sensitive to narrative context dynamically represent information over varying timescales and how these neural patterns align with model-derived features. We also explore how increasing video duration impacts both model and brain representations in tasks requiring multi-scene integration, character motivation inference, narrative summarization, and event boundary detection. Finally, we assess the effect of fine-tuning MLLMs on movie-specific narrative data in enhancing the models’ understanding of characters, emotions, and discourse, and evaluate how these enriched representations correlate with brain activity during movie watching. Our findings aim to elucidate the parallels between human and machine narrative comprehension, informing both cognitive neuroscience and the development of more context-aware AI systems.
Understanding how humans and artificial intelligence systems process complex narrative videos is a fundamental challenge at the intersection of neuroscience and machine learning. We investigate how the temporal context length of video clips (3--24 \,s clips) and the narrative-task prompting shape brain-model alignment during naturalistic movie watching. Using fMRI recordings from participants viewing full-length movies, we examine how brain regions sensitive to narrative context dynamically represent information over varying timescales and how these neural patterns align with model-derived features. We find that increasing clip duration substantially improves brain alignment for multimodal large language models (MLLMs), whereas unimodal video models show little to no gain. Further, shorter temporal windows align with perceptual and early language regions, while longer windows preferentially align higher-order integrative regions, mirrored by a layer-to-cortex hierarchy in MLLMs. Finally, experiments with four narrative-task prompts show that they elicit task-specific, region-dependent brain alignment patterns and context-dependent shifts in clip-level tuning in higher-order regions. Our work positions long-form narrative movies as a principled testbed for studying long-timescale temporal integration in long-context MLLMs and its relationship to cortical responses during narrative comprehension.

\end{abstract}

\section{Introduction}
The alignment between internal representations of  Transformer-based models and cortical activation patterns elicited by naturalistic stimuli has emerged as a key focus in the study of brain-model correspondence~\citep{toneva2019interpreting,schrimpf2021neural,caucheteux2020language,oota2025deep,goldstein2025unified}. 
In particular, prior work has demonstrated that contextual representations extracted from such Transformer models outperform static word embeddings, offering improved brain alignment and capturing temporal dynamics across model depth in both text~\citep{toneva2019interpreting} and speech~\citep{vaidya2022self,oota2024speech}. More recently, multimodal Transformer models have been shown to achieve a higher degree of brain predictivity across various stimulus modalities~\citep{nakagi2024brain,subramaniam2024revealing,dong2023interpreting,oota2024multi,sartzetaki2024one}.
However, the role of longer temporal context in brain alignment for naturalistic multimodal stimuli remains underexplored. Therefore, extending these insights to naturalistic narrative movies is important because comprehension depends on integrating information over longer timescales.

\begin{figure}
    \centering
   \includegraphics[width=\linewidth]{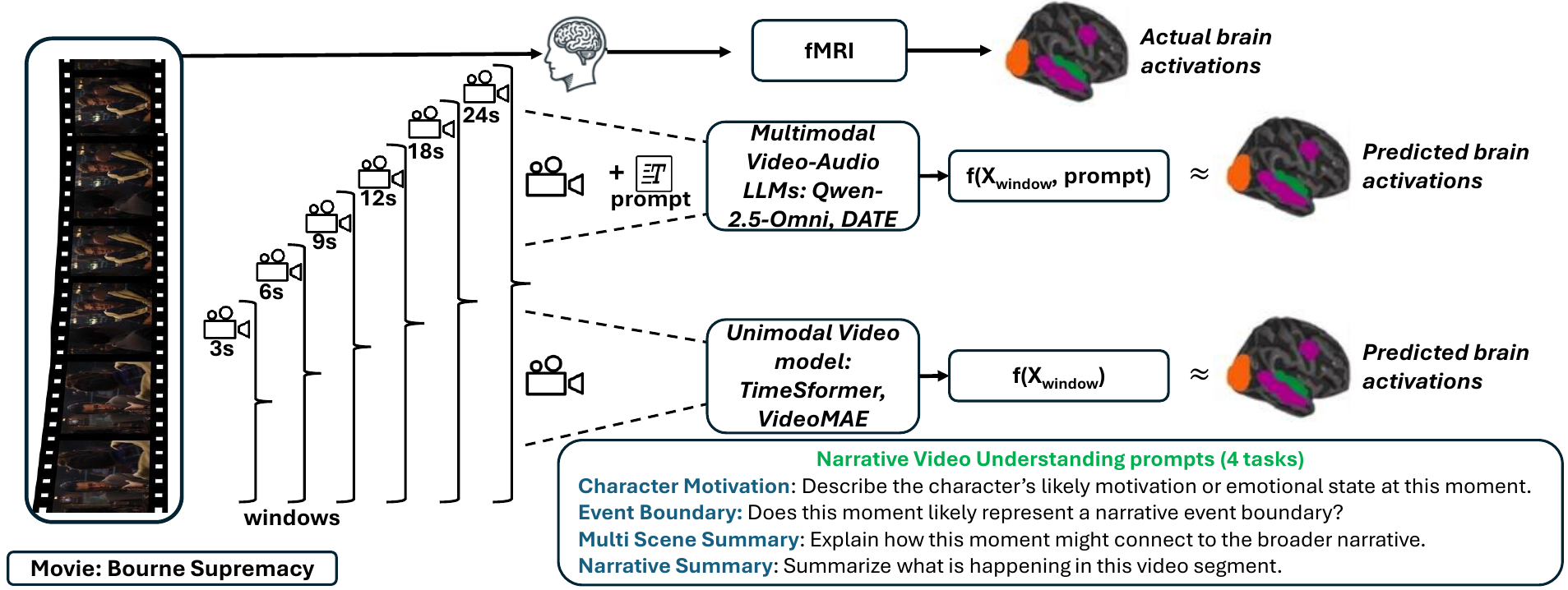}
    \caption{Leveraging temporal video context of different durations ($X_{\text{windows}}$) with unimodal and  multimodal models for brain encoding with a diverse set of instructions (prompts). We experiment with 4 narrative video understanding tasks: character motivation, event boundary detection, multi-scene summarization, narrative summarization.}
    \label{fig:approach}
\end{figure}

Understanding how narrative videos are processed by both AI models and the human brain is an important open problem in cognitive neuroscience and AI. Humans seamlessly integrate visual scenes and the context of the story over time, drawing on sight and sound to comprehend movies. For current AI models, this remains challenging: earlier video Transformer models often processed videos as frame sequences with limited context windows. Processing long sequences of frames are compute intensive and can exceed memory constraints, hence earlier models often resort to subsampling frames or truncating context. Consequently, previous \textit{multimodal} brain-encoding studies have typically relied on short clips (e.g., 1.49\,s), thereby neglecting extended temporal context~\citep{subramaniam2024revealing,dong2023interpreting,oota2024multi,sartzetaki2024one}.
% Prior work has typically examined the role of temporal context only within unimodal stimuli settings, while multimodal studies have often relied on short video clips (e.g., 1.49 seconds) from long movies, thereby neglecting extended temporal context. 

Narrative comprehension is a complex cognitive process that involves the integration of visual, auditory, and contextual information over time. It recruits a hierarchy of cortical regions that operate on multiple temporal scales, from sensory processing of moment-to-moment events to the integration of extended narrative context across scenes and subplots~\citep{lerner2011topographic,simony2016dynamic}. fMRI studies using continuous movie stimuli have shown that higher-order brain areas, including regions within the default mode network, are especially sensitive to longer narrative contexts, often requiring tens of seconds of uninterrupted input to represent story-level meaning~\cite{baldassano2018representation,khosla2021cortical}. This motivates evaluating brain alignment of video-audio models on naturalistic movies while systematically increasing the temporal context window.
% Fortunately, recent multimodal LLMs tuned on videos, can process longer clips-spanning seconds to minutes-capturing narrative elements such as events, interactions, and scene context beyond short snippets. Also, recent multimodal brain encoding studies with narrative videos report that recent instruction-tuned multimodal models can effectively process both video and audio together~\citep{oota2025instruction}. 
% Given the precedence of importance of long context in text and speech, studying its impact for multimodal videos is critical. Hence, in this work, we systematically investigate temporal context from multimodal Transformer models in the presence of rich multimodal stimuli (naturalistic movies with video along with audio). Specifically, we assess how does the length of narrative video context 
% %(temporal context) 
% influence the information extracted by multimodal models, and assess the brain regions which process video information from both short and long temporal context windows. Further, since the best multimodal LLMs (MLLMs) are instruction tuned, we can investigate whether such models, when prompted with narrative understanding tasks, generate task-specific representations that vary with temporal context. Also, how does this variability influence brain alignment? To our knowledge, this is the first study to systematically understand the combined impact of varying clip duration and narrative tasks on brain alignment using such MLLMs.

Using brain recordings of participants watching several popular movies with audio~\citep{Ecole50613}, we investigate the brain alignment of two pretrained video-audio MLLMs. Specifically, as shown in Fig.~\ref{fig:approach}, we evaluate Qwen-2.5-Omni~\citep{xu2025qwen2} and DATE~\citep{yuan2025date}, which are end-to-end multimodal models that jointly process video and audio in an interleaved way, rather than treating them separately. We also experiment with two unimodal video models (TimeSFormer~\citep{gberta_2021_ICML} and VideoMAE~\citep{tong2022videomae}). We evaluate four narrative video understanding tasks and vary the temporal context using clip duration (3s, 6s, 9s, 12s, 18s, and 24s) and estimate brain alignment across tasks and temporal context. Overall, this study addresses the following research questions (RQs): 
\begin{enumerate}[noitemsep, topsep=0pt]
    \item How does increasing the temporal context-length (3-24 s clip duration) affect the brain predictivity of video-audio MLLM and unimodal video model representations during naturalistic movie watching?
    \item Which brain regions show the largest gains (or shifts in optimal window length) in brain predictivity as the temporal context increases, and how are these effects related to the layer-wise representations of video-audio MLLMs?
    \item Which narrative-task instructions (character motivation, event boundary detection, multi-scene summary, and narrative summary) exhibit the highest brain-alignment of MLLM representations during naturalistic movie watching, and do these task-specific representations dissociate into distinct ROI-specific (region of interest) patterns?
    \item Which video clips are most predictive of voxel responses across temporal context lengths and narrative tasks, and how do these patterns vary across ROIs?
\end{enumerate}

Our analysis of longer temporal context and narrative-task prompting in video-audio MLLMs yields several key conclusions corresponding to the RQs: 
(1) Increasing temporal context (clip duration from 3\,s to 24\,s) systematically improves brain predictivity for video-audio MLLMs, while unimodal video models show little to no improvement with longer context windows. (2) We find an ROI-specific temporal gradient: longer temporal windows align best with higher-order semantic regions (e.g., posterior cingulate cortex (PCC)), whereas shorter-to-intermediate windows (3--6\,s) are optimal for perceptual and early language regions (e.g., posterior temporal lobe (PTL)), mirroring a layer-wise hierarchy in MLLMs.
(3) Probing MLLMs with narrative task instructions reveals that \textit{Narrative Summary} and \textit{Multi-scene Summary} explain a larger portion of voxels in higher-order language ROIs, while \textit{Character Motivation} preferentially aligns with more localized temporal language regions; suggesting that task instructions can be used as functional probes of brain-aligned representations.
(4) Interpretation of video clips that most strongly drive voxel responses show that the maximally activating clips are largely the same across temporal windows in visual ROIs, but change with temporal context in higher-order language ROIs.

Overall, these results demonstrate that temporal context and narrative‑task prompting act as complementary probes for long‑form narrative video understanding in the brain, and for interpreting long‑context representations in video-audio MLLMs. We present one of the first systematic evaluations of extended temporal context and narrative‑task prompting in video-audio MLLMs, showing that context‑ and task‑dependent representations are associated with distinct patterns of brain alignment across cortical regions and model layers. 
%The code is part of supplementary material.

% A detailed discussion of related work on contextual representations and brain alignment, and multimodal brain encoding is provided in Appendix~\ref{app:relatedwork}. We summarize representative evaluation settings, particularly stimulus modality and temporal window length in Table~\ref{tab:multimodal_settings}.

\section{Related work}
\label{app:relatedwork}

\noindent\textbf{Contextual Representations and Brain Alignment.}
% Understanding how high-level contextual representations in modern language models map onto human brain activity has become a central thread in cognitive computational neuroscience and NLP models~\citep{jain2018incorporating,toneva2019interpreting,aw2022training,vaidya2022self}.
Our work is most closely related to studies by~\citet{jain2018incorporating,toneva2019interpreting,aw2022training,vaidya2022self,oota2024speech}, who investigate how high-level contextual representations from modern language models map onto human brain activity, and examine how contextual dynamics affect brain-model alignment in language-related ROIs. \citet{jain2018incorporating,toneva2019interpreting,aw2022training} focus on text-based language models with context lengths ranging from 1 to 500 words, whereas~\citet{vaidya2022self,oota2024speech} study speech-based models with context windows ranging from 1.49 to 64 seconds. 
Beyond text and speech, a growing body of work models brain responses during naturalistic movie viewing using representations from unimodal and multimodal models, and more recently from MLLMs, providing strong baselines for video-to-brain encoding~\citep{dong2023interpreting,dong2023vision,oota2025correlating,oota2024multi}. Motivated by these studies, we study how long-range visual context in naturalistic videos affects brain–model alignment, comparing unimodal encoders with MLLMs.

\noindent\textbf{Multimodal Brain Encoding.}
Our work also relates to a growing literature on aligning AI model representations with human brain activity under naturalistic stimuli. 
Several studies have used unimodal video models and multimodal models (e.g., video+audio) to predict movie-evoked brain activity with strong predictive performance~\citep{dong2023interpreting,dong2023vision,oota2024multi}. We summarize representative evaluation settings, particularly stimulus modality and temporal window length in Table~\ref{tab:multimodal_settings}. However, much of this prior multimodal work relies on short clips or limited temporal windows, which can underrepresent the role of extended temporal context in naturalistic video understanding.
Complementary to these efforts, our work builds on and extends this line of work by proposing an architecture-agnostic evaluation protocol to characterize context- and instruction-dependent representational changes across varying time windows and tasks, and relate these changes to brain alignment. 
%----Manish sir updated below one
% Our work aligns with efforts to map AI representations to brain activity under naturalistic stimuli. Prior studies using video and multimodal models show strong predictive performance~\citep{dong2023interpreting,dong2023vision,oota2024multi}, but often rely on short clips that undercapture long-range context. We extend this line by introducing an architecture-agnostic protocol to study how context and instructions shape representations across varying time windows and tasks, and relate to brain alignment.

\setlength{\tabcolsep}{1pt}
\begin{table}[t]
\centering
\scriptsize
%\vspace{-0.2cm}
\caption{Overview of multimodal model evaluation settings in brain encoding studies.}
\label{tab:multimodal_settings}
\resizebox{\textwidth}{!}{ 
\begin{tabular}{|p{2cm}|p{5cm}|p{4cm}|p{1.5cm}|p{2.8cm}|c|}
\hline
\textbf{Study} & \textbf{Model Type} & \textbf{Stimulus Modality} & \textbf{Brain Data} & \textbf{Dataset} &\textbf{Video Len}\\
\hline
\citet{popham2021visual} & Vision-Only CNNs vs. Vision-Language & Unimodal (Silent Videos) & fMRI &  Gallant lab short video clips & 2s\\ \hline
\citet{tang2023brain} & Non-instruction-tuned multimodal model (BridgeTower) & Unimodal (Silent Videos), Unimodal (listening stories) & fMRI &  Gallant lab short video clips & 2s\\
\hline
% \citet{nakagi2024brain} & Language models (BERT, GPT-2, Lllama2, OPT) & Multimodal (Videos with audio) &fMRI & 8.3 hours of video dataset & \ding{55}\\
% \midrule
\citet{subramaniam2024revealing} & Non-instruction-tuned multimodal models (SLIP-CLIP, SimCLR, BLIP, BEiT) & Image frame-text pairs (Movies) & SEEG & AMMT & 4s\\\hline
\citet{lahner2024modeling} & Resnet50 with Temporal Shift Module & Video clips& fMRI&BOLD Moments & 3s \\ 
\hline
\citet{dong2023interpreting} & Non-instruction-tuned multimodal models (MERLOT Reserve) & Multimodal (Movies: Videos with audio) & fMRI & Neuromod Friends dataset & 35s\\
\hline
\citet{oota2024multi} & Non-instruction-tuned multimodal models (TVLT and ImageBind) & Multimodal (Movies: Videos with audio) & fMRI & Neuromod Movie10 & 1.49s\\
\hline
\citet{sartzetaki2024one} &Object and Action recognition models  &Image frames to CNNs, Video clips to video models &fMRI&BOLD Moments & 3s\\ \hline
\citet{oota2025instruction} & Instruction-tuned video and audio MLLMs & Multimodal (Movies: Videos with audio) & fMRI & Neuromod Movie10 & 1.49s\\ \hline
Ours & Unimodal Video models and Video-based MLLMs & Multimodal (Movies: Videos with audio) & fMRI & Neuromod Movie10 & 3s - 24s\\
\hline
\end{tabular}
}
\end{table}

\section{Dataset Curation and Tasks}
\paragraph{Brain imaging dataset.} 
We experiment with Movie10~\citep{Ecole50613}, a multimodal naturalistic fMRI dataset, obtained from the Courtois NeuroMod databank. This dataset was collected while four human subjects (s1, s2, s3, s5; data for s4 and s6 is not public) passively watched four different movies:  \emph{The Bourne supremacy}, \emph{The wolf of wall street}, \emph{Hidden figures} and \emph{Life}. Among these, \emph{Hidden figures} and \emph{Life} are repeated twice, with the repeats used for testing and the remaining movies for training. In this work, we use \emph{Life} movies for testing where we average the two repetitions to reduce noise in brain data.
%This dataset is one of the largest publicly available multimodal fMRI datasets in terms of the number of samples per participant. 
The dataset includes 11,017 TRs (Repetition Time) for training and 2013 TRs for testing. We build encoding models where the train and test sets are totally disjoint. 
%Thus there is no possibility of any information leakage during inference on the test set. 
The fMRI data is collected every 1.49 seconds (= 1 TR). Note that for brain encoding studies, the number of samples per participant is more important than the number of subjects because the predictive models are trained independently for each participant. So, having more samples per participant helps us learn a better predictive model. Further, this is a widely used public dataset in prior work~\citep{oota2024multi,dong2023vision}. 
More details about dataset and preprocessing are in Appendix~\ref{app:detailedsubrois}.

The dataset is already preprocessed and projected onto the surface space (``fsaverage6'').
We use the multimodal parcellation of the human cerebral cortex based on the Glasser Atlas (which consists of 180 regions of interest in each hemisphere) to report the ROI analysis for the brain maps~\citep{glasser2016multi}. 
%This includes four visual processing regions (early visual cortex (EVC), object-related areas (LOC), face-related areas (OFA) and scene-related areas (PPA)), one early auditory area (AC), and eight language-relevant regions, encompassing broader language regions: angular gyrus (AG), anterior temporal lobe (ATL), posterior temporal lobe (PTL), inferior frontal gyrus (IFG), inferior frontal gyrus orbital (IFGOrb), middle frontal gyrus (MFG), posterior cingulate cortex (PCC) and dorsal medium prefrontal cortex (dmPFC), 
We select our language ROIs based on the Fedorenko lab's language parcels~\citep{milton2021parcellation,desai2022proper}.
We show the flatmap with labeled ROIs in Appendix Fig.~\ref{fig:language_flatmap} and list the detailed sub-ROIs of these ROIs in Appendix~\ref{app:detailedsubrois}.

\noindent\textbf{Estimating cross-subject prediction accuracy.}
To account for intrinsic noise in biological measurements, we adapt the method of \citet{schrimpf2021neural,oota2024speech} to estimate the cross-subject prediction accuracy for a model's performance for the Movie10 fMRI dataset. 
Each subject $s$ $\in$ (s1, s2, s3, s5) is chosen as the prediction target and the other three are used to predict this target. %We use a voxel-wise encoding model (see Sec. \ref{sec:modelArch}) to predict one participant's response from others.  The detailed approach is described in Appendix~\ref{app:cross_subject_flatmaps}.
%By subsampling fMRI dataset from four participants, we generate all possible combinations of $s$ participants ($s$ $\in$ [2,4]) for watching movies, and use a voxel-wise encoding model (see Sec. \ref{sec:modelArch}) to predict one participant's response from others.
Note that the estimated cross-subject prediction accuracy is based on the assumption of a perfect model, which might differ from real-world scenarios, yet offers valuable insights into model's performance.
%We estimate cross-subject prediction accuracy by training on the combined brain data from \textit{The Bourne supremacy} and \textit{The wolf of wall street} and testing on the brain data from the movie \textit{Life}.
We present the cross-subject prediction accuracy across voxels for the {Movie10 fMRI} dataset for each of the four participants in Appendix~\ref{app:cross_subject_flatmaps}. %The plots show that across all participants higher activity is observed in the language and visual regions with a max correlation up to 0.4 implying that data has low noise and low cross-subject variability.
%\vspace{-0.2cm}

\noindent\textbf{Narrative Tasks.} 
%\vspace{-0.1cm}
\label{sec:tasks-necessity-sufficiency}
We focus on four core narrative tasks (Character Motivation, Event Boundary Detection, Multi-Scene Summary, and Narrative Summary; see the task instructions in Appendix~\ref{app:narrative_tasks_description}) because together they capture the complementary and irreducible components of narrative understanding. This choice rests on three mild assumptions. First, narrative comprehension can be decomposed into local semantics, agent-centric inference, temporal segmentation, and cross-scene integration \citep{bruner1991narrative,abbott2008cambridge}. Second, a task-conditioned multimodal model is capable of recovering each component \citep{zellers2021merlot,gupta2022dial}. Third, the outputs of these tasks can be linearly, or through a low-dimensional nonlinear readout, composed into downstream predictions \citep{schwartz2022neuroai}. Empirical evidence from neuroimaging studies demonstrate that different cortical regions align with these narrative components, and that combining task outputs improves voxel-level prediction of brain activity during story comprehension \citep{huth2016natural,honey2012slow,wehbe2014simultaneously}.
%\vspace{-0.3cm}

\section{Methodology}
%\textcolor{red}{Add experiments with yet another unimodal model. TimeSformer is a very old model.}
\paragraph{Multimodal Large Language Models (MLLMs).}
%\textcolor{red}{Include Date model and baseline TimeSFormer}
Since humans are capable of simultaneously perceiving the visual and auditory information while watching movies, to investigate long narrative video understanding, we use two pretrained video+text MLLMs, each containing 36 layers: (i) Qwen-2.5-Omni model~\citep{xu2025qwen2}, an end-to-end multimodal model that processes video and audio together in an interleaved way, rather than treating them separately, and (ii) DATE (Dynamic Absolute Time Enhancement) model~\citep{yuan2025date} built on Qwen-2.5-VL model, a plug-and-play framework to improve the capabilities of MLLMs in understanding long videos, particularly for tasks that require precise temporal reasoning and event localization. 
%Details for these models are reported in Table~\ref{neural_models}. 
When prompted with narrative task instructions, these models produce task-specific representations that we use for brain encoding.
%When prompted with natural language narrative-based task instructions, this approach aligns with the way humans process narratives of movies in the brain. These two MLLM models have the advantage of processing both video and audio directly, enabling richer multimodal integration. As a result, the representations obtained from MLLMs provide strong narrative-level movie representations.

% \begin{table}[t]
% \scriptsize
% \centering
% \caption{Unimodal video baselines and video-audio MLLMs}
% \label{neural_models} 
% \begin{tabular}{|l|c|c|c|} 
% \hline
% \textbf{Model Name}& \textbf{MLLM}& \textbf{\#Layers} & \textbf{Modality} \\
% \hline
% Qwen-2.5-Omni&\ding{51}&  36 & Video+Text\\
% DATE&\ding{51}& 36 & Video+Text\\ 
% TimeSFormer&\ding{53} &  12 & Video \\
% VideoMAE&\ding{53} &  12 & Video \\
% \hline
% \end{tabular}  
% \end{table}

% \noindent\textbf{Narrative video understanding tasks.}
% To ensure the diversity of narratives video task-specific instructions while considering video clips as input, we consider 4 instructions, as shown in Table~\ref{prompt_instructions}, and extract the language-guided representations from multimodal models. We considered those tasks as they are generally applicable to any video regardless of the contents in the image frames. 

\noindent\textbf{Extraction of Temporal Context from Long Videos.}
To extract narrative task representations from a MLLM, we use a sliding-window procedure to define temporally coherent segments. We extract consecutive windows of length $W$ seconds with a stride of 1.49\,s, resulting in overlapping temporal segments. Each window was treated as a contiguous video snippet and paired with a narrative task instruction before being passed to the MLLMs. From every segment, the processor samples a window-length-dependent number of frames (9, 9, 12, 16, 24, and 30 frames for $W$ = 3, 6, 9, 12, 18, and 24\,s, respectively), together with synchronized audio to capture richer temporal dynamics. We additionally verify that our findings are robust to this choice via a constant frame-density (1,fps) ablation, reported in Appendix~\ref{app:frame_sampling}.
The model produced hidden state representations across all Transformer layers during token generation; these were averaged across tokens at each layer to yield a compact embedding per layer for each sliding window with respect to task instruction. 
%uniformly sampled 16 frames
%This procedure resulted in temporally grounded, task-aligned representations that explicitly captured narrative information across overlapping windows of the video timeline. 
For the two MLLMs, we use the pretrained Transformer model weights for extracting representations. 
%, which generate hidden state representations at each layer. We then average these hidden state representations at layer level of output generated tokens to obtain final embedding at each layer for each video with respect to task instruction. 

\noindent\textbf{Unimodal Video Baselines.}
We also compare with two unimodal video models, VideoMAE~\citep{tong2022videomae} and TimeSFormer~\citep{gberta_2021_ICML} each with 12 layers. 
%Details for these models are reported in Table~\ref{neural_models}. 
For each video clip, we extract layer-wise hidden state representations using the same sliding-window approach as for MLLMs, but without any task prompts (video-only input). We average these hidden states to obtain a single embedding per layer for each sliding window, and use these embeddings for brain encoding.

%To check the benchmark performance of long narrative video understanding of MLLMs, we use VideoMae~\citep{tong2022videomae} and TimeSFormer~\citep{gberta_2021_ICML} as baseline models which processes both temporal and spatial information in videos. We use the same sliding window approach where the contiguous video snippet is passed to the baseline models. Unlike MLLMs, for baseline models, we input video only in a sliding window fashion without any task instruction. We use the pretrained Transformer weights, which generate hidden state representations at each layer. We then average these hidden state representations to obtain final embedding at each layer for each video across sliding windows. 

% To extract narrative tasks representations from multimodal MLLM for the brain encoding task where considering temporal context from 1.49 to 12 secs, we input a video and task instruction to obtain the embeddings for the language-guided instruction. For multimodal MLLM, we use the pretrained Transformer weights, which generate hidden state representations at each layer. We then average these hidden state representations at layer level of output generated tokens to obtain  final embedding at each layer for each video with respect to task instruction. 

\section{Experiments}
\label{sec:experimental_setup}
\noindent\textbf{Encoding model.}
\label{sec:modelArch}
We train bootstrap ridge regression based voxel-wise encoding models~\citep{deniz2019representation} to predict the fMRI brain activity associated with the stimulus representations obtained from the 4 narrative task-specific instructions for both multimodal LLMs. We employ z-score thresholding separately for both input stimulus representations and brain recordings for training and test datasets. This helps identify and remove extreme outliers that could disproportionately affect the Pearson Correlation results. For each subject, we account for the delay in the hemodynamic response by modeling the hemodynamic response function using a finite response filter (FIR) per voxel with 5 temporal delays (TR) corresponding to $\sim$7.5 seconds~\citep{huth2022gallant}. Formally, at each time step $t$, we encode the stimuli as $X_{t}\in \mathbb{R}^{D}$ and voxels from the brain region $Y_{t}\in \mathbb{R}^{V}$, where $D$ denotes the dimension of the concatenation of 5 delayed TRs and $V$ denotes the number of voxels. 
%Overall, with $N$ such TRs, we obtain $N$ training examples.

\noindent\textbf{Train-test setup.}
We build encoding models where the train and test sets are totally disjoint and the model cannot use any clock relationships from the training data during inference. To be completely clear: independent encoding models are trained for each subject using data concatenated from two movies (11017 TRs). %(\textit{The Bourne supremacy}: 4024 TRs and The \textit{wolf of wall street}: 6993 TRs). 
The test set consisted only of data from the \textit{``Life'' movie} (2028 TRs). Thus, there is no possibility of any information leakage during inference on the test set. Hyper-parameter settings are in Appendix~\ref{app:TrainingDetails}.

\begin{figure}[t]
    \centering
    \includegraphics[width=0.325\linewidth]{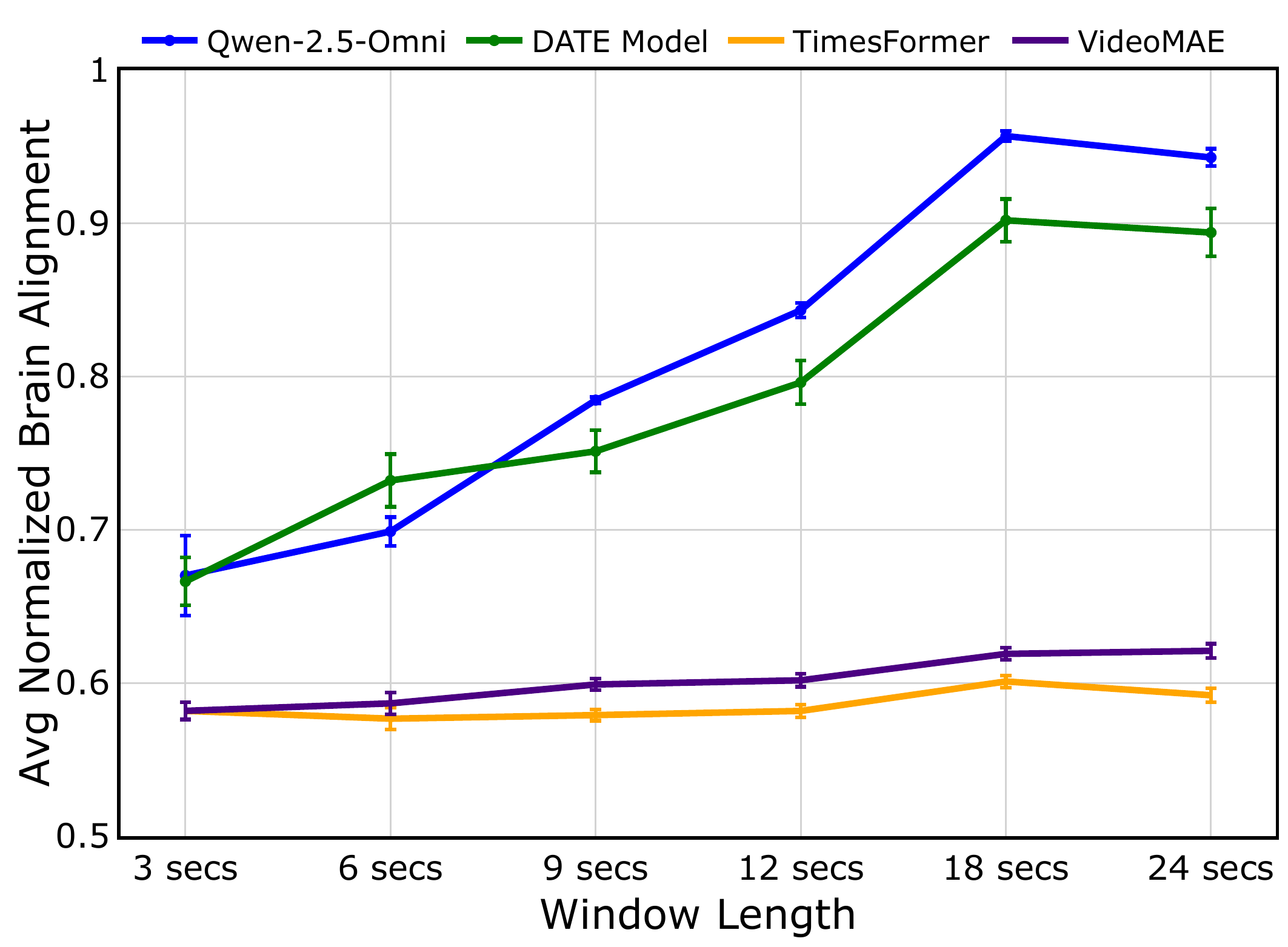}
    \includegraphics[width=0.325\linewidth]{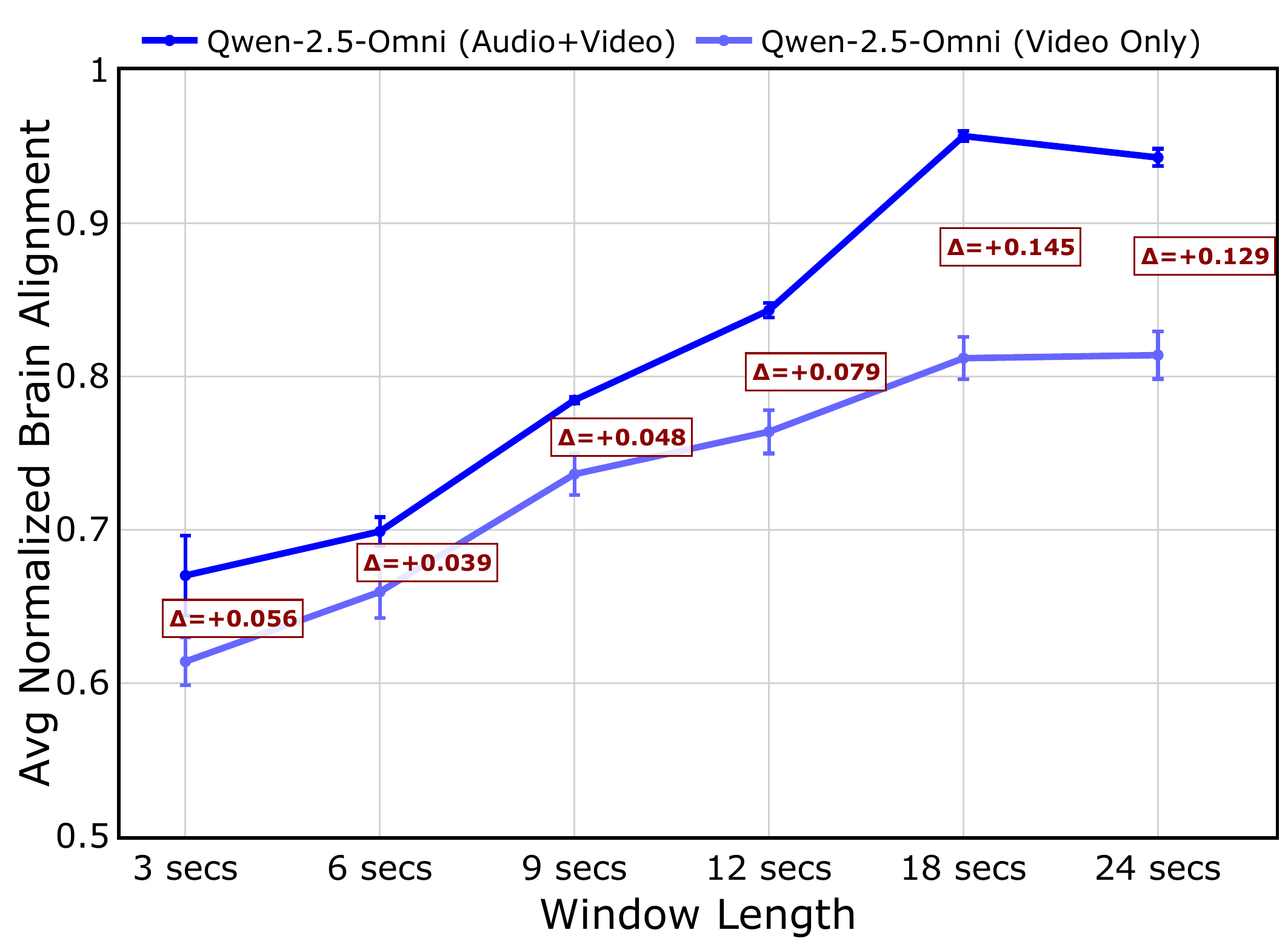}
    \includegraphics[width=0.325\linewidth]{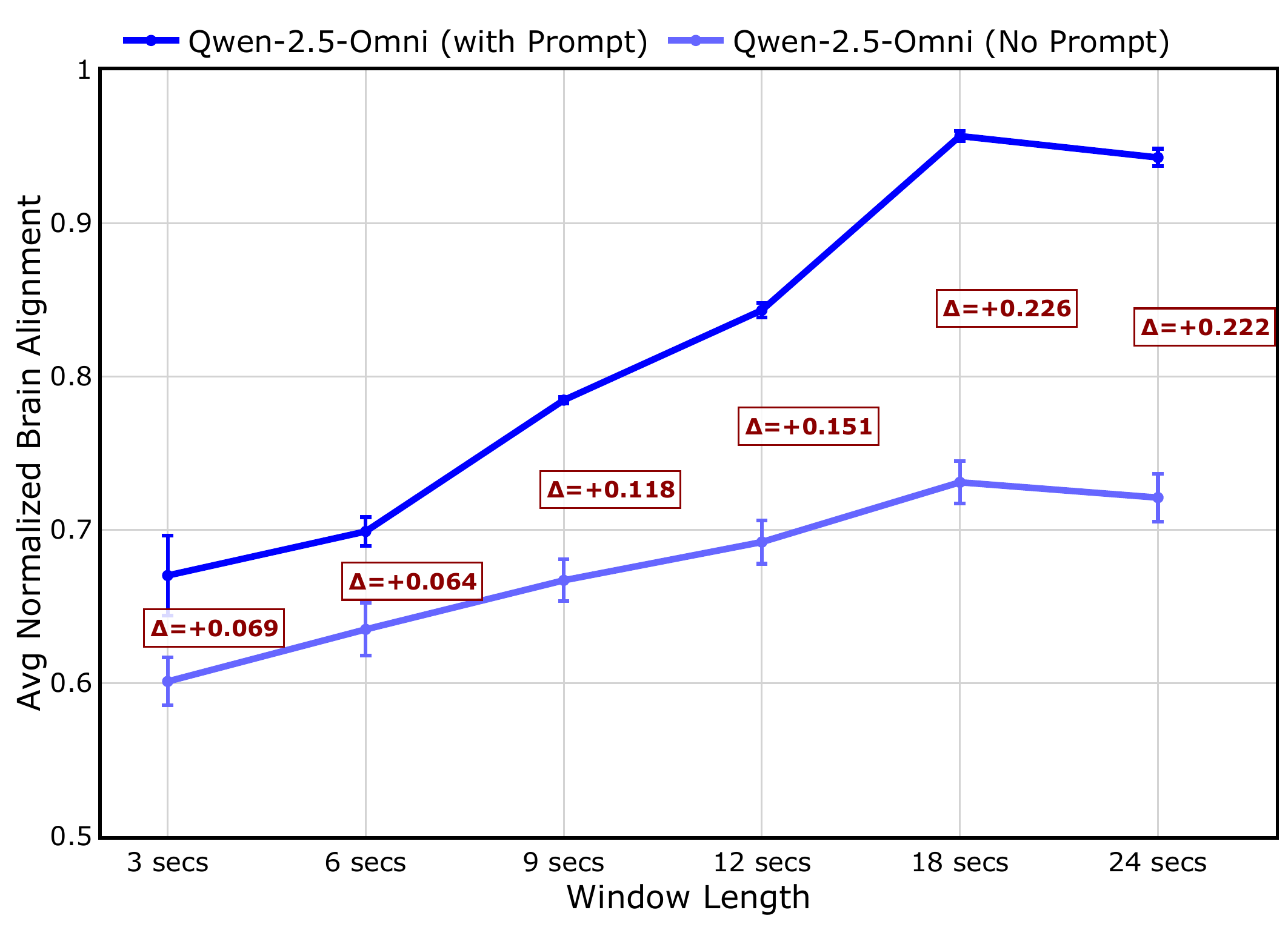}
    %\vspace{-0.2cm}
    \caption{
    % Average normalized brain alignment as a function of temporal window length (3 to 24s). \textbf{(a) Model comparison (left):} MLLMs show increasing brain alignment with longer windows, whereas unimodal video baselines remain approximately constant. \textbf{(b) Modality ablation (middle):} Holding the model fixed (Qwen-2.5-Omni), Video included with Audio input consistently outperforms Video-only input, with the gap widening at longer windows ($\Delta = +0.056$ at 3s to $+0.129$ at 24s). \textbf{(c) Prompting ablation (right):} Holding the model and modality fixed (Qwen-2.5-Omni, Video included with audio), narrative-task prompting yields substantially larger gains over the no-prompt baseline ($\Delta = +0.069$ at 3s to $+0.226$ at 18s). Panels (b) and (c) disentangle the contributions of modality and prompting from temporal context, showing that all three factors contribute additively, with prompting being dominant factor at longer windows. Error bars denote mean$\pm$SEM across subjects.
    Average normalized brain alignment as a function of temporal window length (3 to 24s). (a) Model comparison (left), (b) Modality ablation with Qwen-2.5-Omni (middle), (c) Prompting ablation with Qwen-2.5-Omni and ``Video included with audio'' setting (right). Panels (b) and (c) disentangle the contributions of modality and prompting from temporal context, showing that all three factors contribute additively, with prompting being dominant factor at longer windows. Error bars denote mean$\pm$SEM across subjects.
    }
    \label{fig:baseline_multimodal_windowlength}
\end{figure}

\noindent\textbf{Evaluation metrics.}
We evaluate our models using Pearson Correlation (PC), which is a standard metric to evaluate brain alignment \citep{jain2018incorporating,schrimpf2021neural,goldstein2022shared}. Let TR be the number of time repetitions in the test set. Let $Y=\{Y_i\}_{i=1}^{TR}$ and $\hat{Y}=\{\hat{Y}_i\}_{i=1}^{TR}$ denote the actual and predicted value vectors for a single voxel, respectively. Thus, $Y$ and $\hat{Y}~\in \mathbb{R}^{TR}$. 
%and also $\hat{Y}\in \mathbb{R}^{TR}$. 
% \noindent\textbf{Normalized brain alignment.}
PC is then computed as correlation between the model's predictions $\hat{Y}$ and neural recordings $Y$. 
To quantify the model predictions, the resulting model prediction correlations are divided by the estimated cross-subject prediction accuracy; and averaged across voxels, regions, and participants, resulting in a standardized measure of performance referred to as normalized brain alignment. To calculate \emph{normalized alignment}, we select voxels with cross-subject prediction accuracy $\ge$ 0.05, in line with previous works~\citep{popham2021visual,la2022feature}.

\section{Results}
\label{sec:results}
\subsection*{\textbf{[A1]:} Longer time context boosts brain predictivity for MLLMs but not for unimodal models.}
\noindent\textbf{(a) Avg normalized brain alignment vs. temporal context length.}
First, we quantify brain predictivity on the Movie10 dataset across the temporal context using representations extracted from video-audio MLLMs and unimodal video baselines (TimeSFormer and VideoMAE), as shown in Fig.~\ref{fig:baseline_multimodal_windowlength}a. For each model, we report the average normalized brain alignment across subjects at the best performing layer (layer 36 for MLLMs and layer 12 for unimodal baselines). 

From Fig.~\ref{fig:baseline_multimodal_windowlength}a, we make the following observations: (i) As context increases from 3\,s to 24\,s, both Qwen-2.5-Omni ($\sim$41\% relative gain; 0.67 $\rightarrow$ 0.943) and DATE ($\sim$33.4\% relative gain; 0.67 $\rightarrow$ 0.894) exhibit a consistent increase in brain alignment, whereas unimodal video baselines show little to no change across window lengths (0.58 $\rightarrow$ 0.621).
Paired two-sided \(t\)-tests across subjects (\(df=3\)) further show that Qwen-2.5-Omni significantly outperforms the unimodal video baselines at longer contexts (6s: \(p=0.005\); 9s: \(p=8.5\times 10^{-4}\); 12s: \(p=0.0012\), 18s: \(p=0.00028\); 24s: \(p=0.00024\)), while the difference at 3s is not significant (\(p=0.107\)); DATE shows the same pattern.
In contrast, Qwen-2.5-Omni and DATE do not differ significantly at any window length (all \(p \ge 0.435\)).
Overall, these results suggest that models with integrated video-audio processing gain increased brain alignment from extended temporal context than video-only baselines.

\noindent\textbf{(b) Modality ablation: Video included with Audio vs. Video only.}
To disentangle the contribution of audio modality from temporal context, we hold the model fixed (Qwen-2.5-Omni) and compare ``Video included with Audio'' vs ``Video-only'' input across window lengths (Fig.~\ref{fig:baseline_multimodal_windowlength}b). Audio+Video consistently outperforms Video-only input across all window lengths, with the gap widening from $\Delta$=+0.056 at 3s to $\Delta$=+0.129 at 24s. This indicates that audio provides complementary information beyond what video alone captures, and this complementarity grows with longer temporal context.

\noindent\textbf{(c) Prompting ablation: With prompt vs. no prompt.}
To further disentangle the contribution of narrative-task prompting, we hold the model and modality fixed (Qwen-2.5-Omni, Audio+Video) and compare with-prompt versus no-prompt conditions (Fig.~\ref{fig:baseline_multimodal_windowlength}c). Prompting yields substantially larger gains than the modality contribution, with the gap increasing from $\Delta$=+0.069 at 3s to $\Delta$=+0.226 at 18s. This indicates that narrative-task instructions serve as effective probes that guide MLLM representations toward brain-relevant features, with the largest benefit at longer temporal contexts.

\noindent\textbf{Frame-sampling ablation: constant frame density vs. scaled sampling.} 
A natural concern is whether the gains with longer context arise from temporal span itself or from the increased frame count in our default scaled-sampling scheme. We re-ran the analysis with a constant frame density of 1,fps (Table~\ref{tab:fps_ablation}, Appendix~\ref{app:frame_sampling}). Brain alignment increases monotonically with temporal span under both sampling strategies, with small absolute differences ($\sim$0.03–0.06), confirming that the gains are driven by temporal span rather than by frame density.

\begin{figure*}[t]
    \centering
    \includegraphics[width=0.47\linewidth]{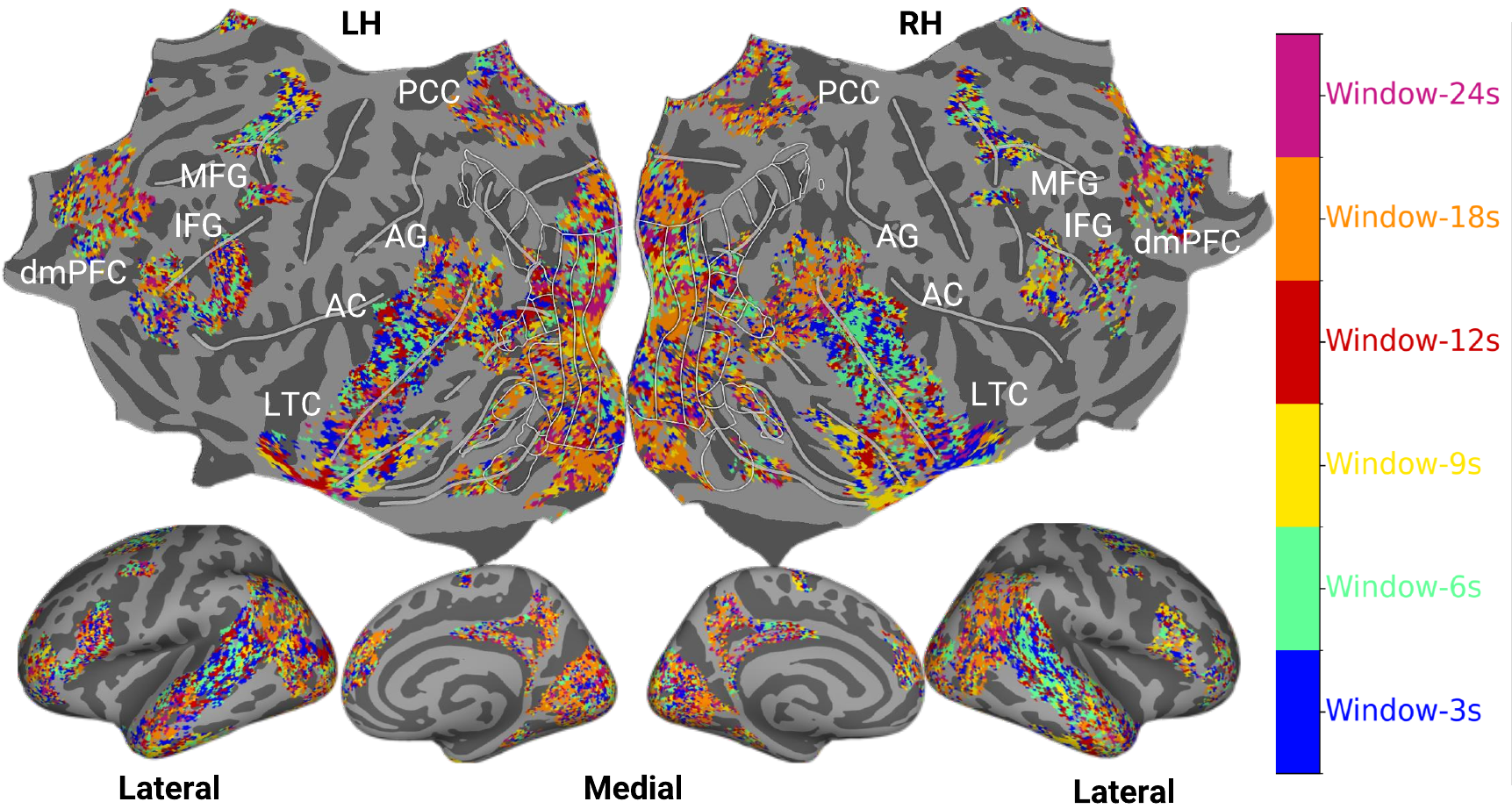}
    \includegraphics[width=0.51\linewidth]{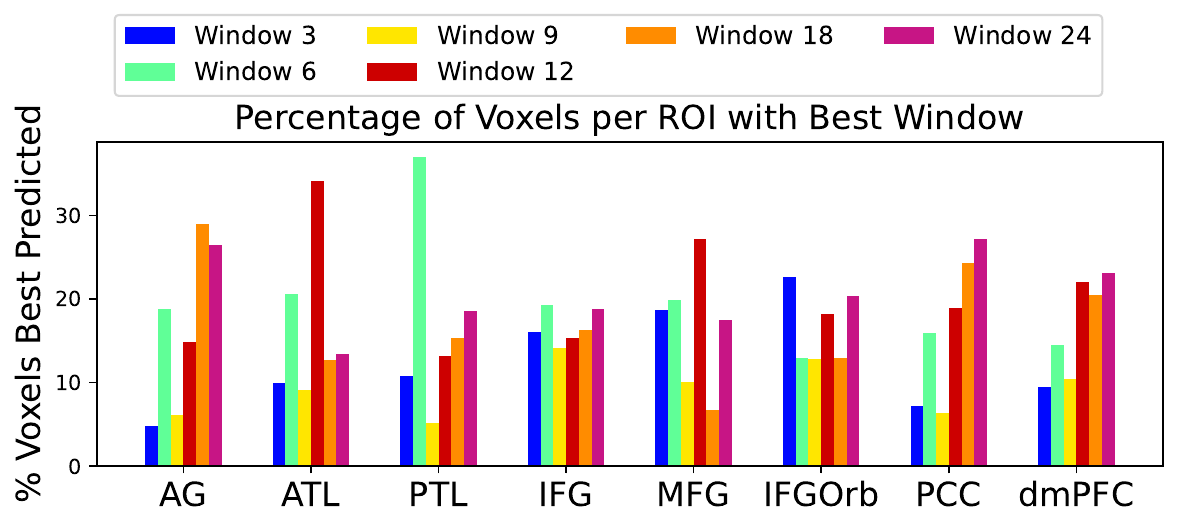}
    %\vspace{-0.2cm}
    \caption{Each voxel is color-coded with the video duration (window length) that led to the highest normalized brain alignment with the Qwen-2.5-Omni model. 
    %The color bar highlights color codes for each window length. 
    (Left) The voxels are projected onto the flattened cortical surface of the `fsaverage' subject.  
    % , with applied hex color codes for the 10 task instructions, 
     (Right): Percentage of best predicted voxels whose brain encoding performance is higher corresponding to each window length within language-selective regions, and visual regions. Results for DATE are in Appendix~\ref{app:window-wise-alignment-date-timesformer}.}    \label{fig:temporal_context_mllm_qwen2.5omni}
\end{figure*}

\subsection*{\textbf{[A2](a)}: Long windows preferentially benefit higher-order semantic regions, while intermediate windows are optimal for several mid-level temporal ROIs.} 
% \noindent\textbf{Temporal context and ROI analysis.}
%To investigate whether temporal context (duration of video clips) is effective in predicting brain activity, and how shorter vs. longer windows impact the brain alignment across language, visual, and auditory regions, we analyze the voxels as follows. 
%Voxel-wise temporal window preference maps (Fig.~\ref{fig:temporal_context_mllm_qwen2.5omni}) (left) reveal an ROI-specific temporal gradient. 
We next examine how shorter vs.\ longer temporal windows differentially affect brain alignment across language, visual, and auditory regions using a voxel-wise window-preference analysis.
Specifically, for each voxel, we select the temporal context (clip duration) that results in the highest normalized brain alignment (averaged across subjects) and apply the window-specific color code to the voxel in Fig.~\ref{fig:temporal_context_mllm_qwen2.5omni} (left).  
We show brain maps for Qwen-2.5-Omni across the six temporal windows.

% Each voxel is color-coded by the temporal context length that maximizes normalized brain alignment (averaged across subjects) and projected onto the flattened cortical surface of the `fsaverage' subject; the corresponding color scheme is shown alongside the maps.
%for average normalized brain predictivity across subjects where the voxel color codes are projected onto the flattened cortical surface of the `fsaverage' subject. The color-scheme corresponding to each temporal window is also reported. 
We observe the following: 
(i) In language ROIs, posterior temporal lobe (PTL) show a clear preference for intermediate context (6\,s), consistent with its role in local syntactic processing over shorter timescales~\citep{friederici2003role,friederici2012cortical,lerner2011topographic}; (ii) Higher-order language ROIs (anterior temporal lobe (ATL), and middle frontal gyrus (MFG) show a strong brain alignment preference for longer context (12\,s), consistent with their roles in semantic composition and lexical processing~\citep{blank2020no,toneva2019interpreting}; (iii) Higher-order narrative associative regions (angular gyrus (AG), posterior cingulate cortex (PCC) and dorsomedial prefrontal cortex (dmPFC)) show a strong brain alignment preference for longer context (18-24\,s), consistent with their roles in narrative integration, episodic memory, and discourse-level processing~\citep{hasson2008hierarchy,lerner2011topographic}; (iv) Frontal regions inferior frontal gyrus (IFG) and orbital IFG (IFGOrb) shows strong alignment for both shorter and longer context, suggesting sensitivity across multiple timescales, consistent with their role in hierarchical syntactic and semantic processing~\citep{friederici2012cortical}; (v) Across ROIs, 9\,s rarely emerges as the dominant window. This does not imply weak performance, 9\,s is often competitive but frequently outperformed by 12\,s for many voxels, causing 12\,s to dominate the winner-take-all map.
Overall, this hierarchical pattern, shorter contexts preferred in lower-order regions and longer contexts in higher-order regions, directly mirrors the known cortical hierarchy of temporal receptive windows~\citep{hasson2008hierarchy,lerner2011topographic}, providing neuroanatomically grounded support for the observed temporal integration gradient.

Fig.~\ref{fig:temporal_context_mllm_qwen2.5omni} (right) quantifies these effects ROI-wise across language network. We make the following observations (i) Most higher-level language ROIs show a plurality of voxels preferring 18\,s and 24\,s, including AG, PCC and dmPFC, which are known to process narratives while ATL and MFG prefers 12\,s, known for lexical information processing.  In contrast, a smaller subset shows a strong intermediate-timescale preference: PTL is  best predicted at window 6s, and ATL and AG also leans toward 6\,s. Notably, 9\,s rarely emerges as the plurality winner across ROIs, suggesting that the dominant transitions are from short (3\,s) to moderate (6\,s) and from moderate (6\,s) to long (12\,s, 18\,s and 24\,s), rather than a smooth monotonic shift. 
% Fig.~\ref{fig:temporal_context_mllm_qwen2.5omni} (right) quantifies these voxel-wise preferences across language ROIs. Higher-order ROIs (ATL, IFG/MFG/IFGOrb, PCC, dmPFC) show a plurality preference for 12\,s, whereas PTL (and AG) prefer intermediate context (6\,s). Across ROIs, 9\,s rarely emerges as the plurality winner, indicating a coarse shift from intermediate (6\,s) to long (12\,s) timescales rather than a smooth progression.

Overall, these results support heterogeneous temporal integration across cortex: long windows preferentially benefit higher-order semantic regions, while intermediate windows are optimal for several mid-level temporal ROIs. Together, these results reveal a temporal gradient: shorter windows emphasize local perceptual-linguistic encoding, whereas longer windows capture abstract narrative integration across distributed language and association networks.

% Overall, these results reveal an ROI-specific temporal integration gradient: intermediate windows (6\,s) are optimal for mid-level temporal language regions (PTL/AG), whereas longer windows (12\,s) are preferred in higher-order association regions (PCC/dmPFC), consistent with longer-timescale narrative integration.

\begin{figure*}[t]
    \centering
    \includegraphics[width=0.6\linewidth]{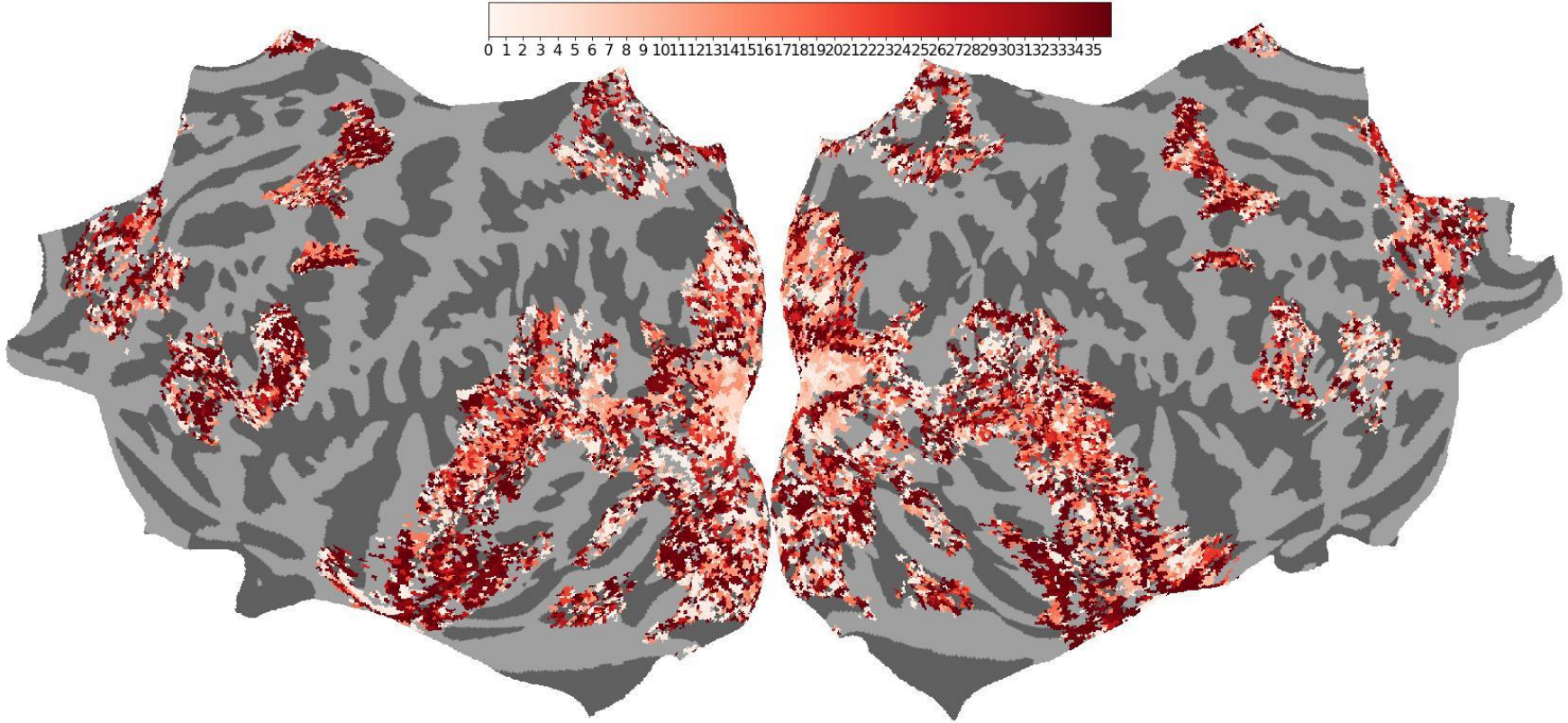}
    \includegraphics[width=0.39\linewidth]{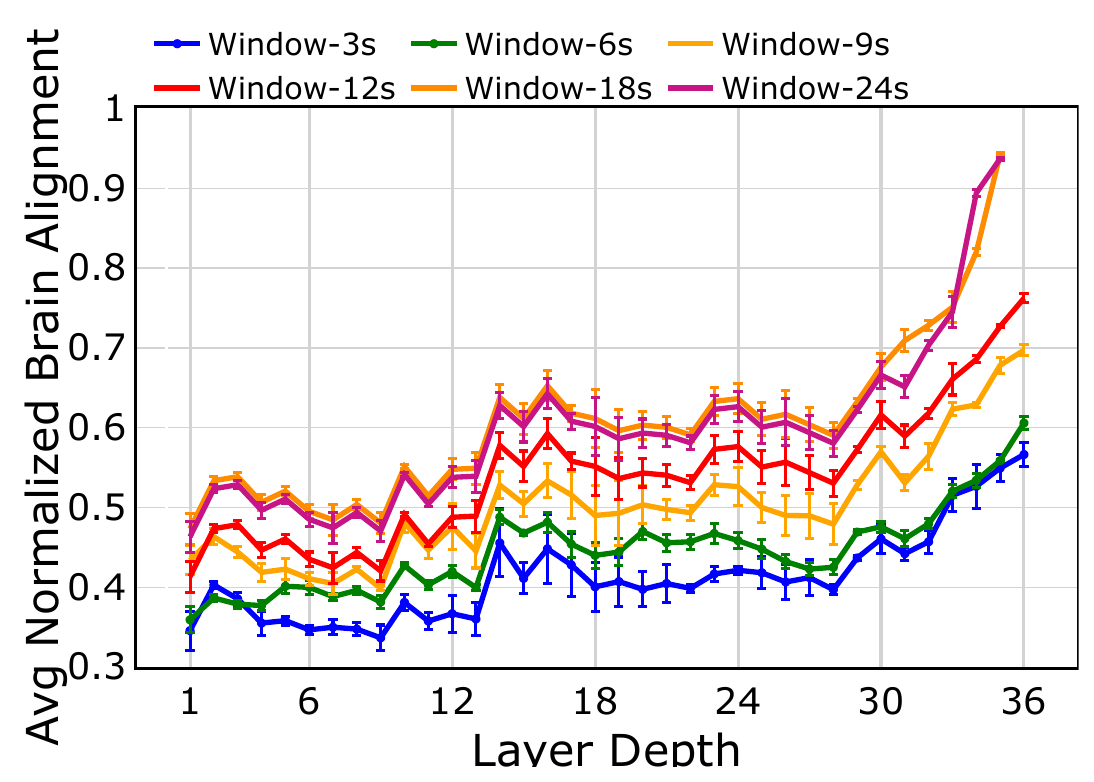}
    %\vspace{-0.2cm}
    \caption{Layer-wise alignment results for Qwen-2.5-Omni: Each voxel is color coded with the MLLM layer number (out of 36) that led to the highest normalized brain alignment for 12\,s setting. The color bar highlights color codes for each layer. 
    The voxels are projected onto the flattened cortical surface of the `fsaverage' subject. Results for DATE are in Appendix~\ref{app:layer-wise-date-timesformer}.}
    \label{fig:qwen2.5-omni_layers}
\end{figure*}

Consistent with the pattern in Fig.~\ref{fig:temporal_context_mllm_qwen2.5omni}, the DATE model relative to Qwen-2.5-Omni, we quantitatively observe that the percentage of voxels in the ROIs shows a more even split between intermediate windows (6-9 s) across several ROIs, while 12 s remains the dominant best window across most ROIs (notably PCC/dmPFC) (see Appendix~\ref{app:window-wise-alignment-date-timesformer} and Fig.~\ref{fig:temporal_context_mllm_date}). 
In contrast, the unimodal video baselines show little to no gain in average normalized brain alignment as temporal context increases. Appendix~\ref{app:window-wise-alignment-date-timesformer} Fig.~\ref{fig:temporal_context_timesformer} shows the brainmap for the unimodal baselines. The voxel-wise maps indicate that longer temporal windows tend to yield higher alignment in higher-order language/association regions, whereas shorter windows are more predictive in early sensory regions.
%We observed similar findings for DATE model, as discussed in Appendix~\ref{app:task-wise-alignment-date-timesformer} in Fig.~\ref{fig:temporal_context_mllm_date}.

%\textcolor{red}{Include some discussion about how these results compare with our baseline model.}

\subsection*{\textbf{[A2](b):} Layer-wise MLLM representations form a cortical language hierarchy and shift with temporal context.}

We investigate how MLLM brain-encoding performance varies across model layers as the temporal context window increases. Across all temporal windows, Fig.~\ref{fig:qwen2.5-omni_layers} (right) shows a consistent stratification into three layer groups: early layers (1--12), middle layers (13--28), and late layers (29--36). Normalized brain alignment generally improves from early to late layers across sliding time windows, suggesting progressively more abstract representations in later layers.

% We investigate how MLLM brain-encoding performance varies across model layers as the temporal context window increases, as shown in Fig.~\ref{fig:qwen2.5-omni_layers} (right). Across all temporal windows, we observe a consistent stratification into three layer groups: early layers (1--12), middle layers (13--28), and late layers (29--36). 
% %Overall, normalized brain alignment increases with longer sliding-window context, indicating that MLLMs benefit from long-range temporal information compared to shorter windows (e.g., 3\,s and 6\,s). 
% We find a layer-wise hierarchy: normalized brain alignment generally improves from early to deeper layers across sliding time windows, suggesting progressively more abstract representations in later layers.

    Fig.~\ref{fig:qwen2.5-omni_layers} (left) visualizes voxel-wise layer preferences for the 12\,s condition, projected onto the \texttt{fsaverage} subject. We observe that (i) early sensory regions (early visual and auditory cortex) align best with lower layers, consistent with shallow representations capturing low-level sensory features; (ii) higher-level visual regions such as lateral occipital complex (LOC) and parahippocampal place area (PPA), and temporal/parietal language ROIs (PTL and AG) tend to align with middle-to-late layers; and (iii) language-related regions such as inferior frontal gyrus (IFG), anterior temporal lobe (ATL), and angular gyrus align most strongly with the deepest layers. Together, these results suggest that MLLMs exhibit a layered representational hierarchy that broadly mirrors cortical processing hierarchies. Layer-wise brain maps for each subject for Qwen-2.5-Omni are shown in Appendix~\ref{app:layer-wise-date-timesformer} Fig.~\ref{fig:qwen_omni_layers_instruction_brainmap}. We observe a similar language hierarchy for the DATE model in Appendix~\ref{app:layer-wise-date-timesformer} Fig.~\ref{fig:qwen2.5-date_layers}.

\begin{figure*}[t]
    \centering
    \includegraphics[width=0.46\linewidth]{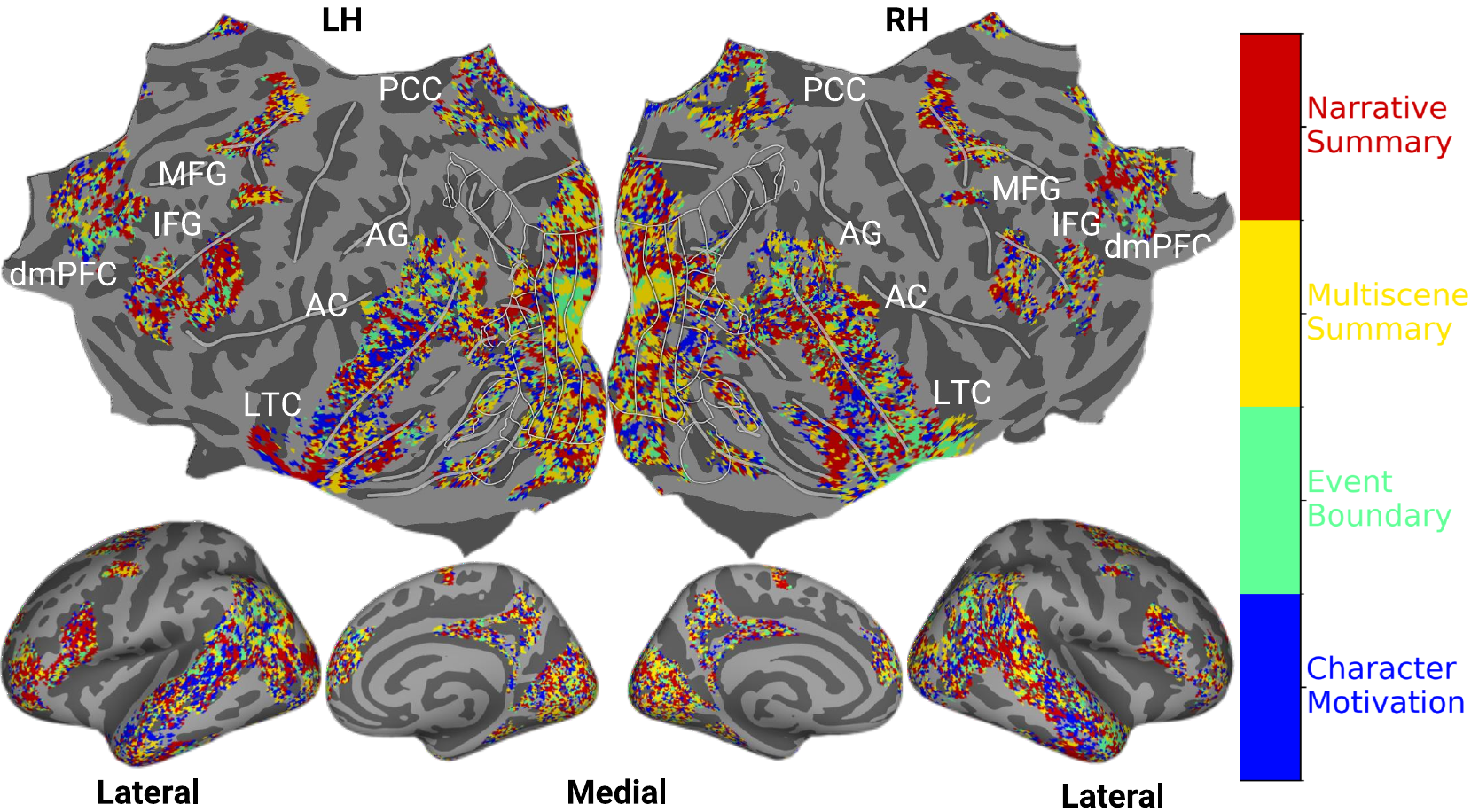}
    \includegraphics[width=0.53\linewidth]{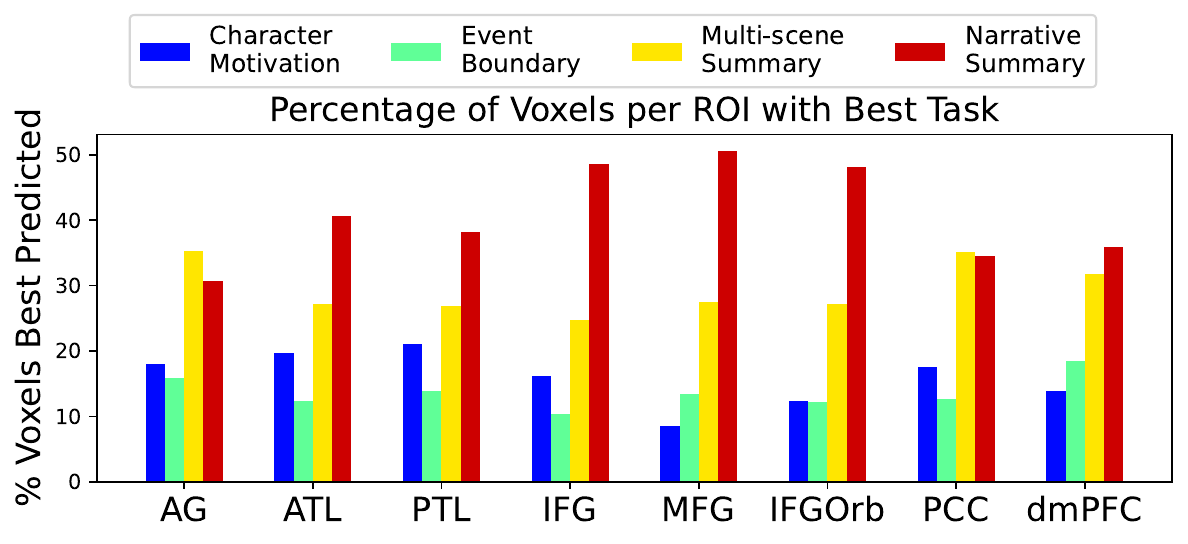}
    %\vspace{-0.2cm}
    \caption{(Left) Each voxel is color-coded with the instruction that led to the highest normalized brain alignment for Qwen-2.5-Omni model. The color bar highlights color codes for each instruction. The voxels are projected onto the flattened cortical surface of the `fsaverage' subject. (Right) Percentage of voxels in each ROI corresponding to each instruction. Results for DATE are in Appendix~\ref{app:task-wise-alignment-date-timesformer}.}
    \label{fig:qwen2.5-omni_tasks}
\end{figure*}

\subsection*{\textbf{[A3]:} Narrative task-specific representations show ROI-specific dissociations while encoding.}
To investigate which task instructions are more effective in predicting brain activity and whether MLLMs differentiate task-specific representations and provide clear separation in brain regions, assign each voxel the task with highest normalized brain alignment and color-code accordingly
% for each voxel, we select the task instruction that results in the highest normalized brain alignment and apply the task instruction-specific color code to the voxel 
(Fig.~\ref{fig:qwen2.5-omni_tasks}, left). For each ROI, we report the fraction of voxels where each task is optimal (Fig.~\ref{fig:qwen2.5-omni_tasks}, right).

%shows brain maps for Qwen-2.5-Omni for tasks for average normalized brain predictivity across subjects where the voxel color codes are projected onto the flattened cortical surface of the `fsaverage' subject. The color-scheme corresponding to each task instruction is also reported. 
From Fig.~\ref{fig:qwen2.5-omni_tasks} (left): (i) \textit{Narrative and multiscene summary} tasks explain a large fraction of voxels across widespread association cortex, consistent with these tasks emphasizing long-range semantic integration over extended narrative context. (ii) \textit{Character Motivation} dominates lateral/ventral temporal and occipito-temporal cortex, suggesting that character-centric inference aligns more strongly with regions supporting person- and situation-level representations grounded in audiovisual content. (iii) \textit{Event Detection} is sparse and localized, indicating that boundary/event-focused representations explain fewer voxels overall and may be confined to more specialized subregions. (iv) The strong intermixing of winning tasks within broad territories suggests that task instruction prompts act as functional probes, revealing heterogeneous, task-selective subpopulations of voxels even within the same large-scale networks.

Fig.~\ref{fig:qwen2.5-omni_tasks} (right) summarizes task preferences across language ROIs. \textit{Narrative Summary} has highest fractional of voxels in frontal and temporal ROIs such as IFG, MFG, IFGOrb, ATL, PTL and in dmPFC, indicating that global narrative integration prompts yield representations most predictive of these regions. \textit{Multi-scene Summary} is especially prominent in integrative hubs such as AG and PCC, where it is comparable to \textit{Narrative Summary}. \textit{Character Motivation} contributes a relatively larger share in temporal language ROIs (notably ATL and PTL), whereas \textit{Event Boundary Detection} is rarely the plurality winner, with comparatively higher contributions in dmPFC (and to a lesser extent AG). Overall, integration-focused prompts (Narrative/Multi-scene Summary) dominate higher-order ROIs, while agent-centric prompts (Character Motivation) contribute more in temporal language areas. We observed similar findings for DATE model (Appendix~\ref{app:task-wise-alignment-date-timesformer} Fig.~\ref{fig:tasks_mllm_date}).
Further, interpretability analyses of voxel-wise encoding weights across tasks and context windows are in Appendix~\ref{app:Interpretability-Analysis-date-timesformer} Fig.~\ref{fig:similarity_windows_tasks}.

% Fig.~\ref{fig:qwen2.5-omni_tasks} (right) display a consistent hierarchy emerges across language ROIs. \textit{Narrative Summary} dominates most frontal and temporal lobe language regions, reaching roughly half of voxels in IFG, MFG, IFGOrb, ATL and PTL, and remaining the plurality winner in dmPFC, indicating that global narrative integration prompts best capture representations predictive of these regions. \textit{Multi-scene Summary} is the second strongest driver and is particularly prominent in integrative hubs such as AG and PCC (where it is comparable to Narrative Summary), consistent with these areas supporting cross-event/scene integration. In contrast, \textit{Character Motivation} accounts for a larger share of best-predicted voxels in more ``local'' language regions (notably ATL and PTL), suggesting that character-centric intent inference aligns better with semantic and discourse-level processing that does not always require full narrative summarization. Finally, \textit{Event Boundary} is rarely the plurality winner, but shows relatively higher contributions in dmPFC (and to a lesser extent AG), consistent with boundary processing being more specialized and distributed. Overall, tasks that explicitly require \emph{long-range semantic integration} (Narrative/Multi-scene Summary) best explain voxels in higher-order language ROIs, whereas more \emph{localized inferential prompts} (Character Motivation) contribute more in temporal language areas.

\noindent\textbf{Robustness of task-instruction prompts.}
To examine whether our task prompts are sensitive to wording or reflect broader task-conditioned representational structure, we performed two additional analyses using MLLM-derived task-conditioned embeddings: (i) paraphrased rewrites preserving meaning, and (ii) generic prompts stripped of task-specific content (see Appendix~\ref{app:narrative_tasks_description} Table~\ref{prompt_instructions_rephrased}). We computed cosine similarity between the original prompt-conditioned embeddings and those obtained under the rephrased and generic prompts from MLLMs. Original-to-rephrased similarities are uniformly high ($>$0.93), while original-to-generic similarities drop substantially (0.67-0.75).
% \begin{table}[t]
% \centering
% \caption{Semantic similarity between original task instructions and paraphrased versus generic rewrites.}
% \label{tab:prompt_rephrased}
% \begin{tabular}{|l|c|c|}
% \hline
% Task & Original$\leftrightarrow$
% Rephrased & Original$\leftrightarrow$
% Generic \\
% \hline
% Character Motivation & 0.934 & 0.738 \\
% Event Boundary & 0.939 & 0.674 \\
% Multi-Scene Summary & 0.943 & 0.754 \\
% Narrative Summary & 0.949 & 0.743 \\
% \hline
% \end{tabular}
% \end{table}
This confirms that our task prompts encode meaningful task-specific semantic structure that is preserved under paraphrasing but lost when content is genericized, supporting the interpretation that observed brain-alignment patterns reflect genuine task-conditioning rather than surface text effects.

\begin{figure*}[t]
    \centering
    \includegraphics[width=\linewidth]{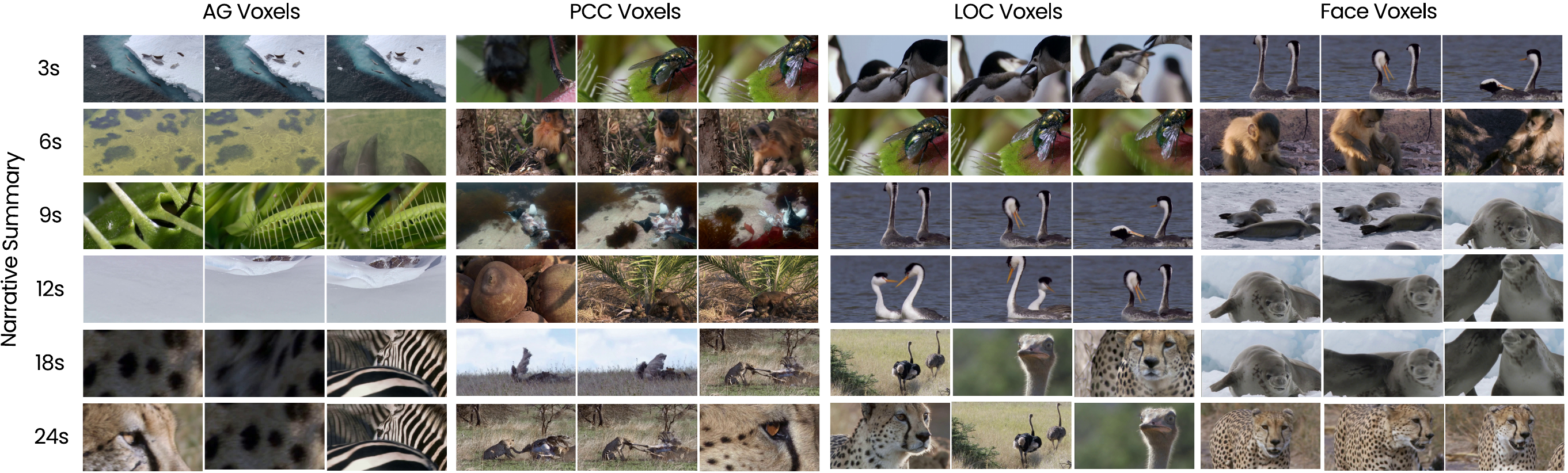}
    %\vspace{-0.2cm}
    \caption{Temporal window conditioned top-K video clips driving voxel responses for Narrative summary task: Results for ROI selective voxels for Subject-1. We identify the top-10 videos from the test stimulus with highest average activation in each temporal window of ROI selective voxels
for the AG, PCC, LOC and Face regions. \emph{For visualization, we show only the single highest-ranked video segment in each condition; the images are sampled frames from within that W-second segment}.
%The image display sample frames from top video that show maximum activation for ROI specific voxels.
}
    \label{fig:qwen2.5-omni_video_specific_voxels}
\end{figure*}

%\vspace{-0.2cm}
\subsection*{\textbf{[A4]:} Maximally activating video clips are stable in visual ROIs but shift with temporal context in higher-order language ROIs.}

To select the video clips that elicit the highest response in a given voxel, we used the voxelwise estimated encoding-model weights across temporal windows. For each temporal window \(w\) and voxel \(v\), let \(\beta^{(w)}_{v}\in\mathbb{R}^{F}\) denote the learned weight vector and \(x^{(w)}_{c}\in\mathbb{R}^{F}\) denote the feature vector for clip \(c\). We scored every clip by the dot product $s^{(w)}_{v}(c)=(\beta^{(w)}_{v})^\top x^{(w)}_{c}$, and selected largest scoring \(K\)=10 clips. 
% These top-ranked clips are those that the encoding model predicts will elicit the strongest responses in voxel \(v\) for window \(w\), providing an interpretable characterization of voxel tuning in the model feature space.
Top-ranked clips are those predicted by encoding model to elicit the strongest responses in voxel \(v\) for window \(w\), offering an interpretable view of voxel tuning in the feature space.

\noindent\textbf{Window-conditioned retrieval.}
We analyze 1000 best-predicted voxels across 6 temporal windows to identify which video clips drive neural responses during long-form movie processing. 
Fig.~\ref{fig:qwen2.5-omni_video_specific_voxels} shows retrieved clips for representative voxels in visual (LOC, face-selective cortex) and higher-order language regions (AG, PCC) across context windows. 
We quantify temporal consistency using Jaccard similarity (J) of retrieved videos across window pairs. Visual ROIs show high consistency (LOC: J=0.50; Face: J=0.38) with category-coherence (object-/shape-dominated for LOC and face/head-dominated for face-selective voxels). But, AG (Win3$\rightarrow$Win24: $\Delta$J = 0.094) and PCC (Win3 $\rightarrow$ Win24: $\Delta$J = -0.153) exhibit larger shifts in retrieved clips as context length increases, suggesting greater sensitivity to broader semantic and narrative context.

% In Fig.~\ref{fig:qwen2.5-omni_video_specific_voxels} we visualize, for AG-, PCC-, LOC- (place) and Face- selective voxels, the top-10 out of
% 2013 video clips from the fMRI stimulus set (Life Movie). From Fig.~\ref{fig:qwen2.5-omni_video_specific_voxels}, we make the following observations: (i) For both LOC and Face voxels across all windows, the selected video clips are visually coherent and “object-centric” (clear foreground objects/animals) for LOC voxels and retrieved clips tend to show prominent faces/heads (here, mostly animal faces/heads rather than human faces) for Face voxels. Even as the window changes, the model keeps retrieving clips with strong, structured shapes-consistent with LOC’s role in object form processing and FFA's role in face form processing. This stability suggests these voxels are driven primarily by visual content present in the frame, not by longer narrative context. (ii) For language regions AG and PCC, the retrieved clips vary more across windows (different scene types). This is interesting that for a higher-level region where tuning is less about a single visual category and more about broader semantic/contextual structure. 

% Overall, maximally activating video clip retrieval reveals a clear dissociation across cortex: category-selective visual ROIs (LOC and face-selective cortex) show stable across windows, whereas higher-level regions (AG and PCC) exhibit more window-dependent exemplar shifts, consistent with greater sensitivity to semantic integration and longer-range narrative context.

\begin{figure*}[t]
    \centering
    \includegraphics[width=\linewidth]{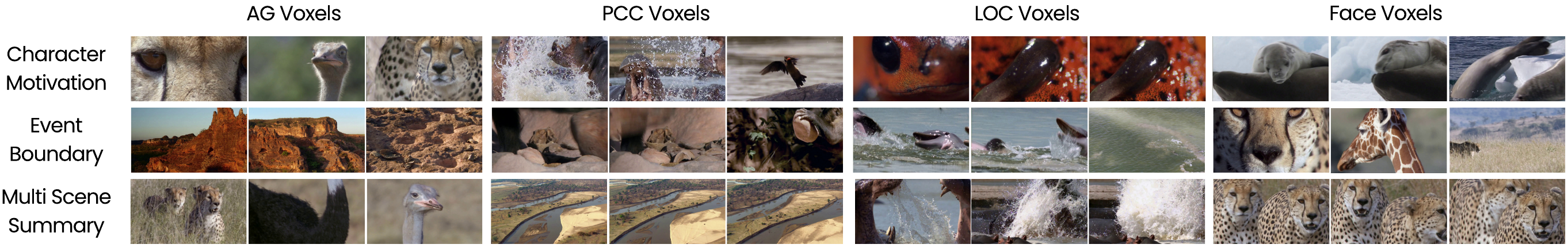}
    \caption{Task conditioned top-K video clips driving voxel responses for longer temporal window (18\,s) for other tasks; 
    %: Results for ROI selective voxels for Subject-1. We identify the top-10 videos from the test stimulus with highest average activation in each temporal window of ROI selective voxels
% for the AG, PCC, LOC and Face regions. \emph{For visualization, we show only the single highest-ranked video segment in each task; the images are sampled frames from within that segment}. R
results for narrative summary task are in Fig.~\ref{fig:qwen2.5-omni_video_specific_voxels}.} 
    \label{fig:qwen2.5-omni_video_specific_voxels_tasks}
\end{figure*}

\noindent\textbf{Task-conditioned retrieval.}
% Fig.~\ref{fig:qwen2.5-omni_video_specific_voxels_tasks} shows retrieval under different narrative-task prompts. Visual ROIs remain stable across tasks, whereas AG and PCC vary more across prompts, indicating that task instructions modulate the semantic/contextual aspects of representations that best align with higher-order regions.   
% \noindent\textbf{Task-conditioned maximally activating video clips.}
We next examine whether the identity of the narrative-task instruction changes which video clips are predicted to maximally activate voxels within a given ROI. Using the same scoring-and-retrieval procedure as above, we select the top-\(K\) clips for ROI-selective voxels under each task instruction and visualize the resulting clips for AG, PCC, LOC, and face-selective voxels. Fig.~\ref{fig:qwen2.5-omni_video_specific_voxels_tasks} shows retrieved clips under different narrative-task prompts. To quantify task-dependent modulation, we compute Jaccard similarity between retrieved video sets for each ROI under different task instructions. Similar to Window-conditioned retrieval, visual ROIs remain stable across tasks (LOC: J = 0.42, Face: J = 0.38), whereas AG and PCC vary more across prompts (AG: J = 0.04, PCC: J = 0.13), indicating that task instructions modulate the semantic/contextual aspects of representations that best align with higher-order regions. 

We observe that (i) in category-selective visual ROIs (LOC and face-selective cortex), the retrieved clips remain visually coherent and category-consistent across tasks (e.g., object/texture- and face/animal-head-dominated clips), suggesting that these regions are driven primarily by stable visual content rather than task framing; (ii) in higher-order regions (AG and PCC), the retrieved clips vary more substantially across task instructions, indicating that task prompting modulates the semantic/contextual aspects of the representation that best align with these ROIs. For example, \textit{Character Motivation} tends to retrieve agent-centric clips with salient animals/faces, whereas \textit{Event Boundary} and summarization-style prompts retrieve more context-rich exemplars.

% %For example, \textit{Character Motivation} tends to retrieve agent-centric clips with salient animals/faces, whereas \textit{Event Boundary} and summarization-style prompts retrieve more scene- and context-rich exemplars (Fig.~\ref{fig:qwen2.5-omni_video_specific_voxels_tasks}). Overall, task-conditioned clip retrieval complements the winner-task voxel maps by providing an interpretable stimulus-level view of how prompt-specific representations differentially drive higher-order ROIs.

Together with the window-based retrieval (Fig.~\ref{fig:qwen2.5-omni_video_specific_voxels}), these results suggest that visual ROIs show stable tuning across both temporal contexts and tasks, whereas higher-order ROIs exhibit stronger task- and context-dependent shifts in the stimuli predicted to drive voxel responses.

\widowpenalty0

\section{Conclusion}
%\vspace{-0.2cm}
We studied how temporal context and narrative-task prompting affect brain–model alignment during naturalistic movie watching using voxel-wise encoding on Movie10, comparing two video–audio MLLMs with unimodal baselines, and obtained these findings.
% We studied how temporal context and narrative-task prompting influence brain-model alignment during naturalistic movie watching. Using voxel-wise encoding on Movie10 fMRI dataset, we compared two video-audio MLLMs against unimodal baselines and obtained these findings. 
\textbf{First}, increasing clip duration (3--24\,s) substantially improves brain alignment for video-audio MLLMs but not for unimodal models (Fig.~\ref{fig:baseline_multimodal_windowlength}). This suggests gains arise from temporal and multimodal integration rather than simply adding frames, highlighting long-form narrative video as a probe of brain-relevant timescales in MLLMs.
\textbf{Second}, context-length preferences reveal an ROI-specific temporal gradient: shorter windows (3--6\,s) align with perceptual and early language regions (PTL), whereas longer windows align with higher-order integrative regions (PCC, dmPFC, ATL); this pattern is mirrored in model depth, with deeper layers better predicting higher-level cortex (Figs.~\ref{fig:temporal_context_mllm_qwen2.5omni}, \ref{fig:qwen2.5-omni_layers}), indicating a representational hierarchy paralleling cortical processing during movie watching.
\textbf{Third}, narrative-task instructions produce task-specific, ROI-dependent representations rather than a single generic embedding: different instructions yield distinct voxel/ROI alignment patterns (Fig.~\ref{fig:qwen2.5-omni_tasks}), and the top-ranked video clips driving voxels differ by ROI and context (Fig.~\ref{fig:qwen2.5-omni_video_specific_voxels}, \ref{fig:qwen2.5-omni_video_specific_voxels_tasks}). 
\textbf{Fourth}, visual ROIs (LOC, face-selective cortex) show stable clip preferences across windows, while higher-order regions (AG, PCC) exhibit strong context-dependent shifts.
Overall, video-audio MLLMs capture a hierarchy and temporal integration that track cortical processing during narrative processing. Task conditioning provides a useful functional probe of brain-aligned representations. Our results bridge neuroscience studies of long-timescale narrative processing with interpretable evaluations of long context MLLMs. We discuss limitations in
Appendix~\ref{app:limitation}.

% To our knowledge, this is one of the first systematic work to link temporal context length and narrative-task prompting to ROI- and layer-specific brain-model alignment during long-form movie stimuli. More broadly, our results bridge neuroscience studies of long-timescale narrative comprehension with interpretable evaluations of long-form multimodal language models.
% These findings show video-audio MLLMs capture a hierarchical, long-timescale temporal integration that parallels cortical narrative processing; task conditioning provides a useful functional probe of these brain-aligned representations; and together the results bridge neuroscience studies of long-timescale narrative processing with interpretable evaluations of long-form MLLMs.

% \section*{Impact Statement}
% This work offers interpretable insights into timescale- and region-specific correspondences between cortical activity and model representations during naturalistic movie watching. For AI community, this work proposes a neuro-grounded evaluation protocol for long-context multimodal models, useful for diagnosing whether architectural changes (temporal modules, fusion) produce representations that behave more like biological integration. 

% We did not create any new neural recordings data as part of this work. We used the Movie10 dataset which is publicly available without any restrictions. Movie10 dataset can be downloaded from \url{https://github.com/courtois-neuromod/movie10/tree/33a97c01503315e5e09b3ac07c6ccadb8b887dcf}. Please read their terms of use\footnote{\url{https://docs.cneuromod.ca/en/latest/ACCESS.html}} for more details. We do not foresee any harmful uses of this technology. 

\bibliography{iclr2026_conference}
\bibliographystyle{icml2026}

%%%%%%%%%%%%%%%%%%%%%%%%%%%%%%%%%%%%%%%%%%%%%%%%%%%%%%%%%%%%%%%%%%%%%%%%%%%%%%%
%%%%%%%%%%%%%%%%%%%%%%%%%%%%%%%%%%%%%%%%%%%%%%%%%%%%%%%%%%%%%%%%%%%%%%%%%%%%%%%
% APPENDIX
%%%%%%%%%%%%%%%%%%%%%%%%%%%%%%%%%%%%%%%%%%%%%%%%%%%%%%%%%%%%%%%%%%%%%%%%%%%%%%%
%%%%%%%%%%%%%%%%%%%%%%%%%%%%%%%%%%%%%%%%%%%%%%%%%%%%%%%%%%%%%%%%%%%%%%%%%%%%%%%
\appendix
% \onecolumn
\noindent{\Large\textbf{Overview of Appendix Sections}}
\begin{description}[leftmargin=!, labelwidth=2cm, noitemsep, topsep=0pt]
%\item[Appendix~\ref{app:relatedwork}] Related Work
\item[Appendix~\ref{app:detailedsubrois}] Dataset and Detailed sub-ROIs of language, visual and auditory regions
\item[Appendix~\ref{app:cross_subject_flatmaps}] Cross-subject prediction accuracy
\item[Appendix~\ref{app:narrative_tasks_description}] Description of Narrative video understanding tasks
\item[Appendix~\ref{app:TrainingDetails}] Implementation details for reproducibility
\item[Appendix~\ref{app:statSig}] Statistical Significance
\item[Appendix~\ref{app:window-wise-alignment-date-timesformer}] Window-wise Voxel Alignment for DATE and TimeSformer
\item[Appendix~\ref{app:frame_sampling}] Robustness to Frame Sampling.
\item[Appendix~\ref{app:layer-wise-date-timesformer}] Layer-wise Analysis for DATE
\item[Appendix~\ref{app:task-wise-alignment-date-timesformer}] Task-wise Voxel Alignment for DATE
%\item[Appendix~\ref{app:videoClips}] Maximally activated video clips
\item[Appendix~\ref{app:Interpretability-Analysis-date-timesformer}] Interpretability Analysis for Qwen2.5-Omni and DATE
\item[Appendix~\ref{app:limitation}] Limitations
\end{description}

\section{Dataset and Detailed sub-ROIs of language, visual and auditory regions}
\label{app:detailedsubrois}

\noindent\textbf{Brain imaging dataset.} 
We experiment with Movie10~\citep{Ecole50613}, a multimodal naturalistic fMRI dataset, obtained from the Courtois NeuroMod databank. This dataset was collected while four human subjects (s1, s2, s3, s5; data for s4 and s6 is not public) passively watched four different movies:  \emph{The Bourne supremacy ($\sim$100 mins)}, \emph{The wolf of wall street ($\sim$170 mins)}, \emph{Hidden figures ($\sim$120 mins)} and \emph{Life ($\sim$50 mins)}. Among these, \emph{Hidden figures} and \emph{Life} are repeated twice, with the repeats used for testing and the remaining movies for training. In this work, we use \emph{Life} movies for testing where we average the two repetitions to reduce noise in brain data.
This dataset is one of the largest publicly available multimodal fMRI datasets in terms of the number of samples per participant. It includes 4024 TRs (Time Repetitions) 
of \emph{The Bourne supremacy} and 6993 TRs of \emph{The wolf of wall street} for training and 2013 TRs of \emph{Life} as test data. We build encoding models where the train and test sets are totally disjoint. 
%Thus there is no possibility of any information leakage during inference on the test set. 
The fMRI data is collected every 1.49 seconds (= 1 TR).

The dataset is already preprocessed and projected onto the surface space (``fsaverage6'').
We use the multimodal parcellation of the human cerebral cortex based on the Glasser Atlas (which consists of 180 regions of interest in each hemisphere) to report the ROI (region of interest) analysis for the brain maps~\citep{glasser2016multi}. This includes four visual processing regions (early visual cortex (EVC), object-related areas (LOC), face-related areas (OFA) and scene-related areas (PPA)), one early auditory area (AC), and eight language-relevant regions, encompassing broader language regions: angular gyrus (AG), anterior temporal lobe (ATL), posterior temporal lobe (PTL), inferior frontal gyrus (IFG), inferior frontal gyrus orbital (IFGOrb), middle frontal gyrus (MFG), posterior cingulate cortex (PCC) and dorsal medium prefrontal cortex (dmPFC), based on the Fedorenko lab's language parcels~\citep{milton2021parcellation,desai2022proper}.

The data covers seven brain regions of interest (ROIs) in the human brain with the following sub-divisions: (i) early visual (EV: V1, V2, V3, V3B, and V4); (ii) object-related areas (LO1 and LO2); (iii) face-related areas (OFA), (iv) scene-related areas (PPA), (v) middle temporal (MT: MT, MST, LO3, FST and V3CD), (vi) late language regions, encompassing broader language regions: angular gyrus (AG: PFm, PGs, PGi, TPOJ2, TPOJ3), lateral temporal cortex (LTC: STSda, STSva, STGa, TE1a, TE2a, TGv, TGd, A5, STSdp, STSvp, PSL, STV, TPOJ1), inferior frontal gyrus (IFG: 44, 45, IFJa, IFSp) and middle frontal gyrus (MFG: 55b)~\citep{baker2018connectomic,milton2021parcellation,desai2022proper}.

\begin{figure}[!ht]
    \centering
    \includegraphics[width=\linewidth]{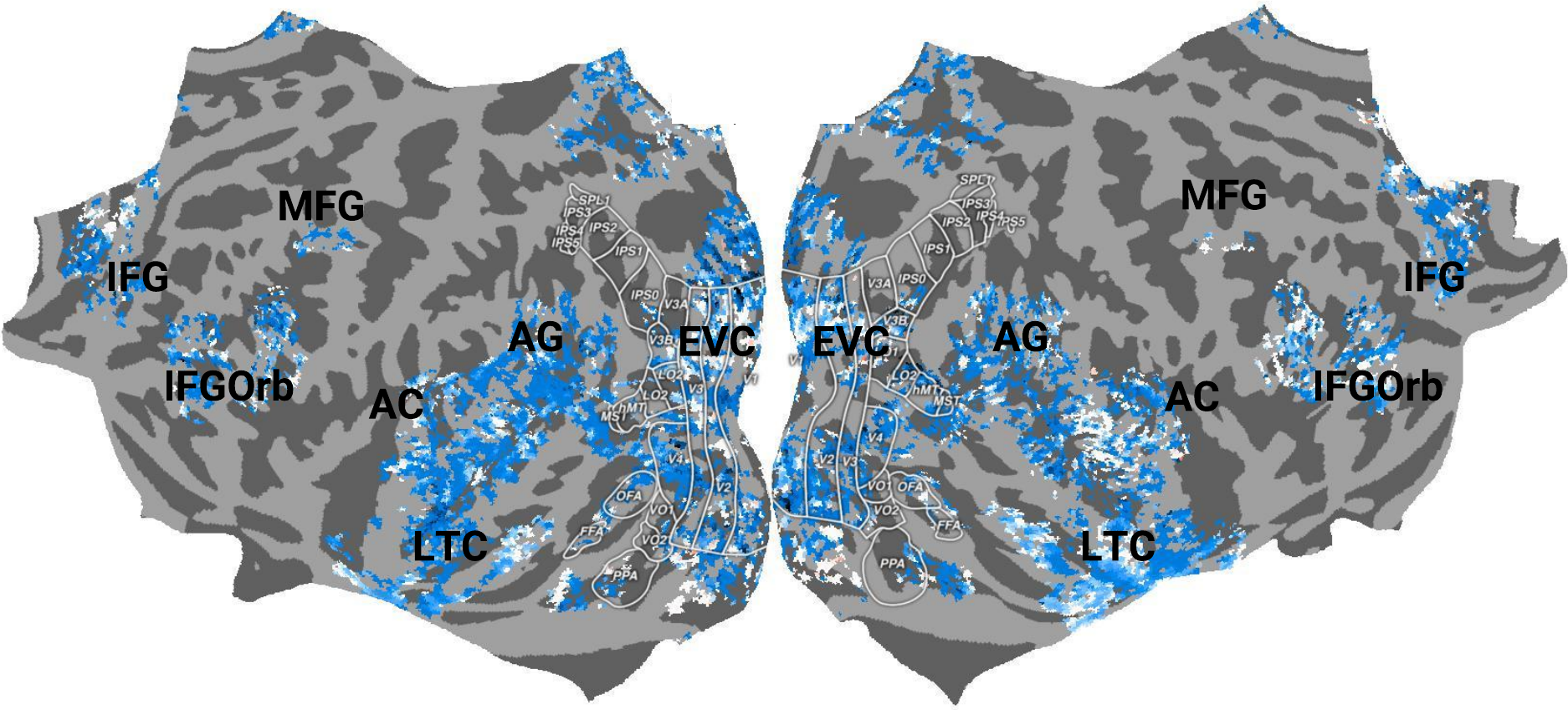}
    \caption{Flattened cortical surfaces for
language-, visual- and auditory-selective regions displayed on the `fsaverage' surface, used as the mask for all participants.}
    \label{fig:language_flatmap}
\end{figure}

\section{Cross-subject prediction accuracy}
\label{app:cross_subject_flatmaps}

We follow the method introduced by~\citet{schrimpf2021neural} to estimate how well brain activity in one individual can be predicted from others, using the Movie10 fMRI dataset. Starting with data from $n$ participants (e.g., $n=4$), for each subject $s$ $\in$ (s1, s2, s3, s5) is chosen as the prediction target and the other three are used to predict this target, we use a voxel-wise encoding model (see Sec. \ref{sec:modelArch}) to predict one participant's response from others. For every combination, one participant was randomly chosen as the target, and the model was trained to predict their brain responses using data from the remaining $n-1$ participants. This gave us an average prediction score (correlation) for each voxel at each participant.
To extrapolate to infinitely many humans and thus to obtain the highest possible (most conservative) estimate, as suggested by~\citet{schrimpf2021neural}, we fit the equation $v=v_0\times \left(1-e^{-\frac{x}{\tau_0}}\right)$ where $x$ is each subsample's number of participants, $v$ is each subsample's correlation score and $v_0$ and $\tau_0$ are the fitted parameters. 
This fitting was performed for each sensor independently with 100 bootstraps each to estimate the variance where each bootstrap draws $x$ and $v$ with replacement. The final ceiling value was the median of the per-voxel ceilings $v_0$.

\begin{figure*}[!ht] 
\centering
\begin{minipage}{\textwidth}
\centering
    \includegraphics[width=0.6\linewidth]{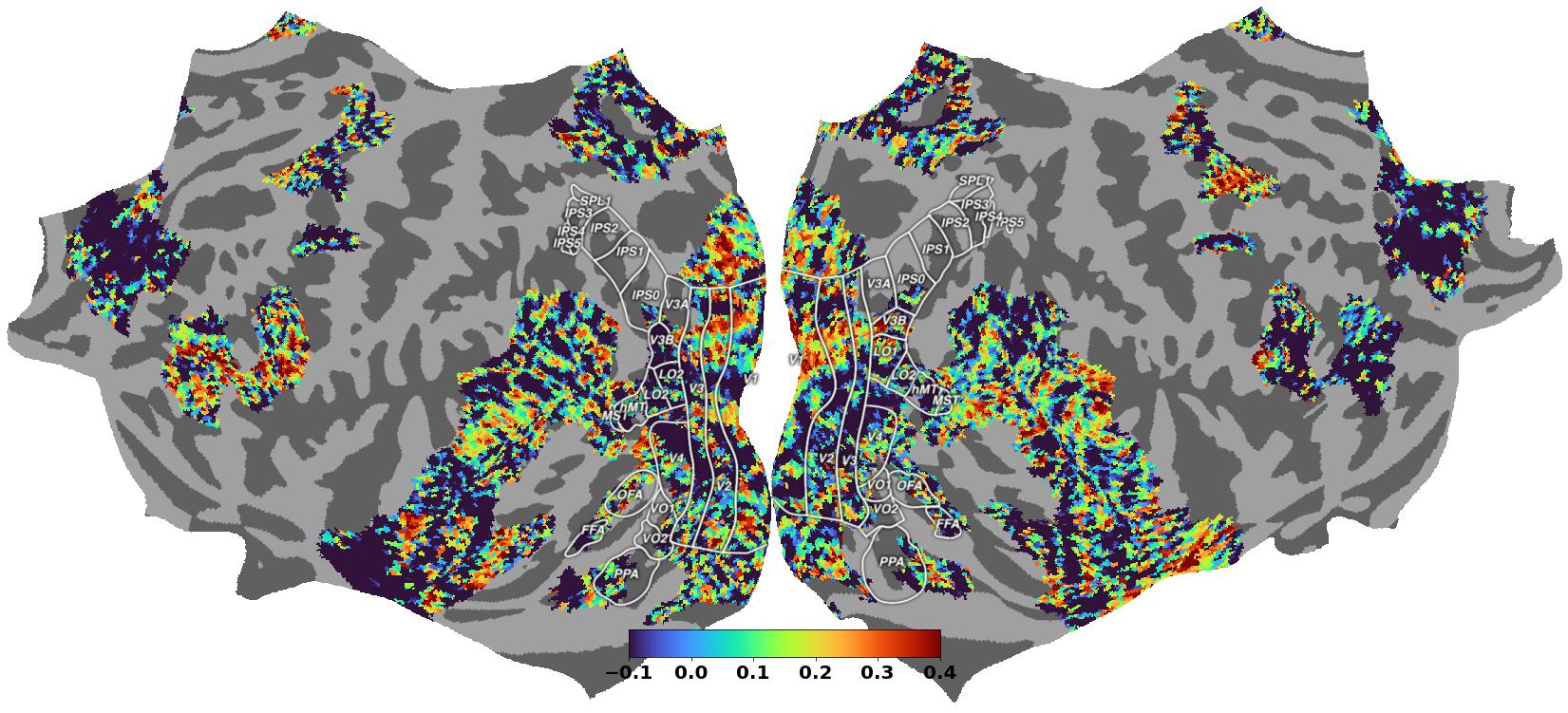}
    \\(a) Subject-01 \\
\end{minipage}
\begin{minipage}{\textwidth}
\centering
    \includegraphics[width=0.6\linewidth]{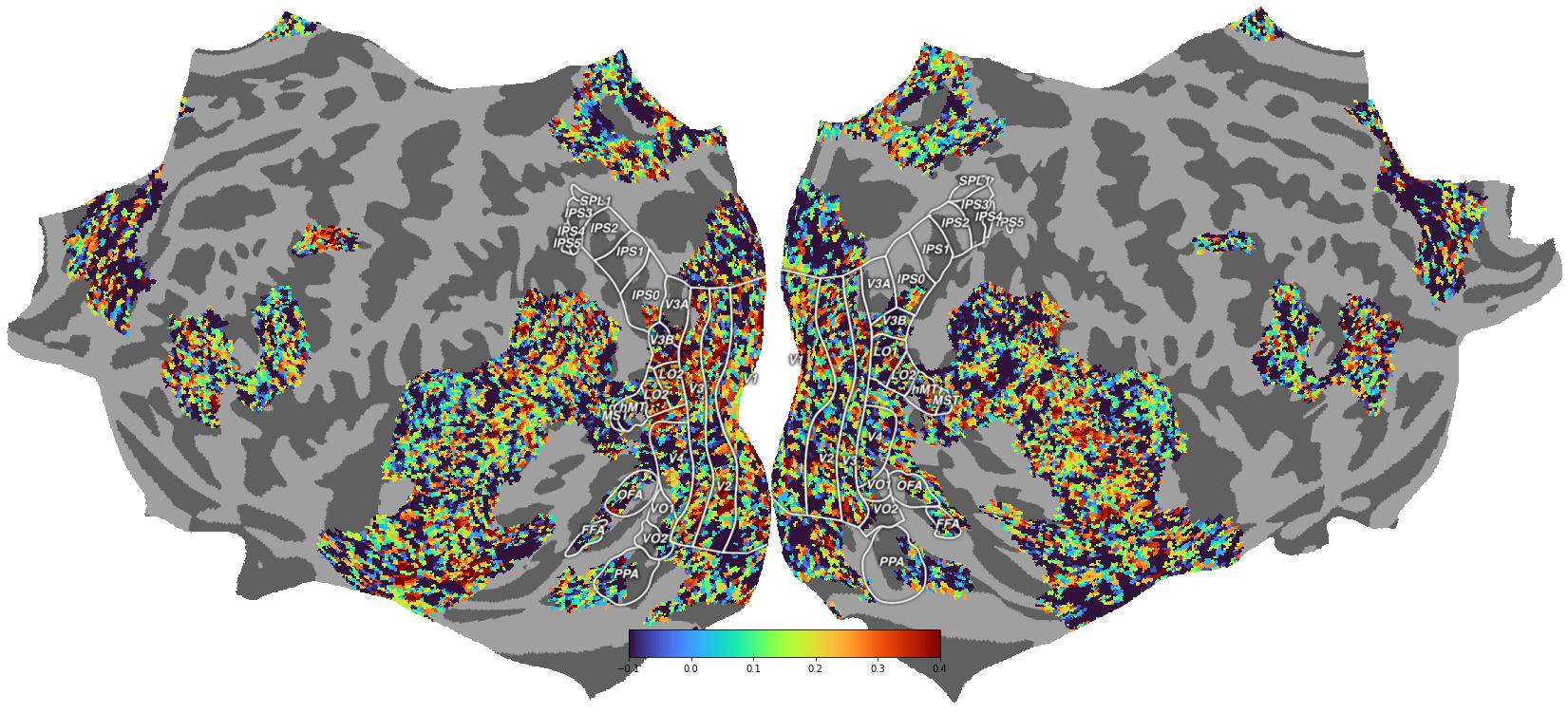}
    \\(a) Subject-02 \\
\end{minipage}
\begin{minipage}{\textwidth}
\centering
    \includegraphics[width=0.6\linewidth]{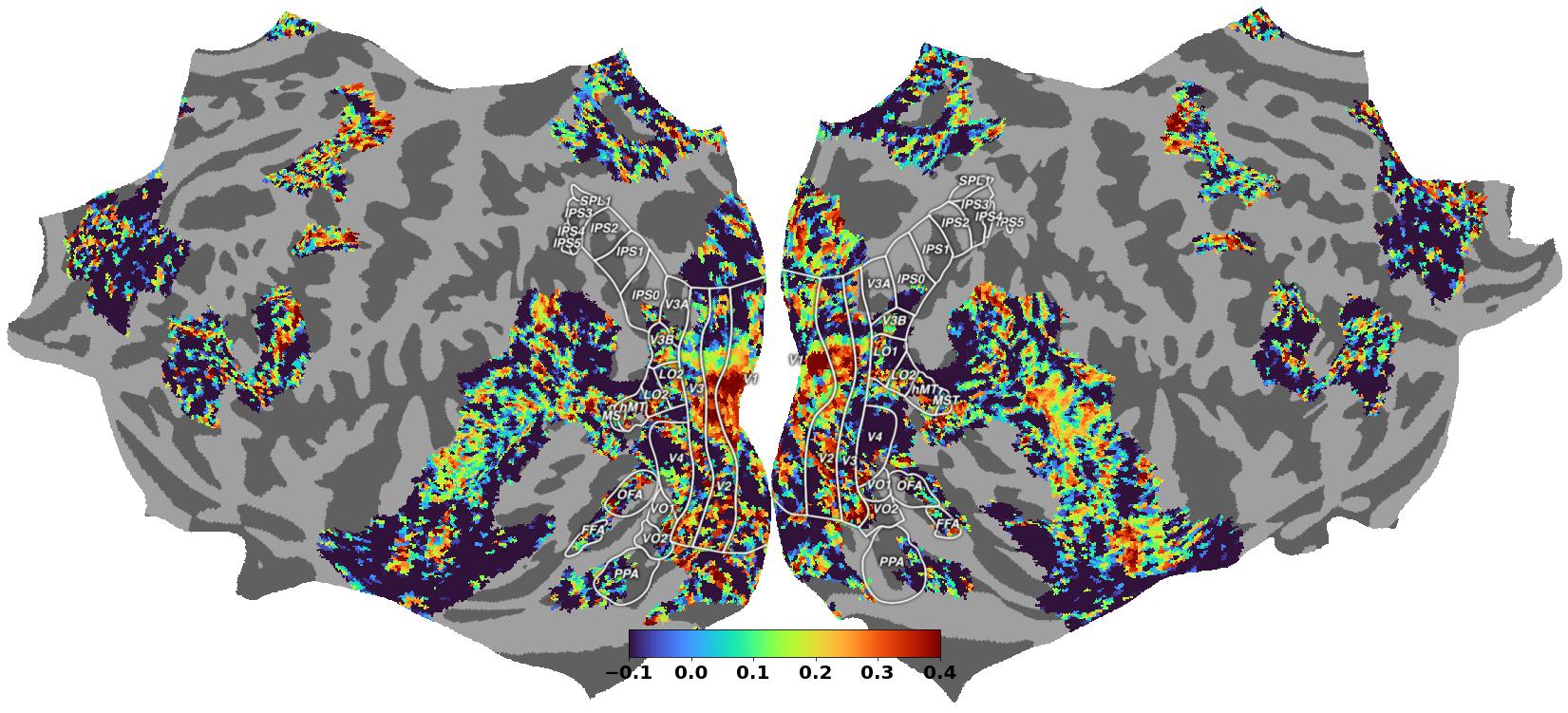}
    \\(b) Subject-03 \\
\end{minipage}
\begin{minipage}{\textwidth}
\centering
    \includegraphics[width=0.6\linewidth]{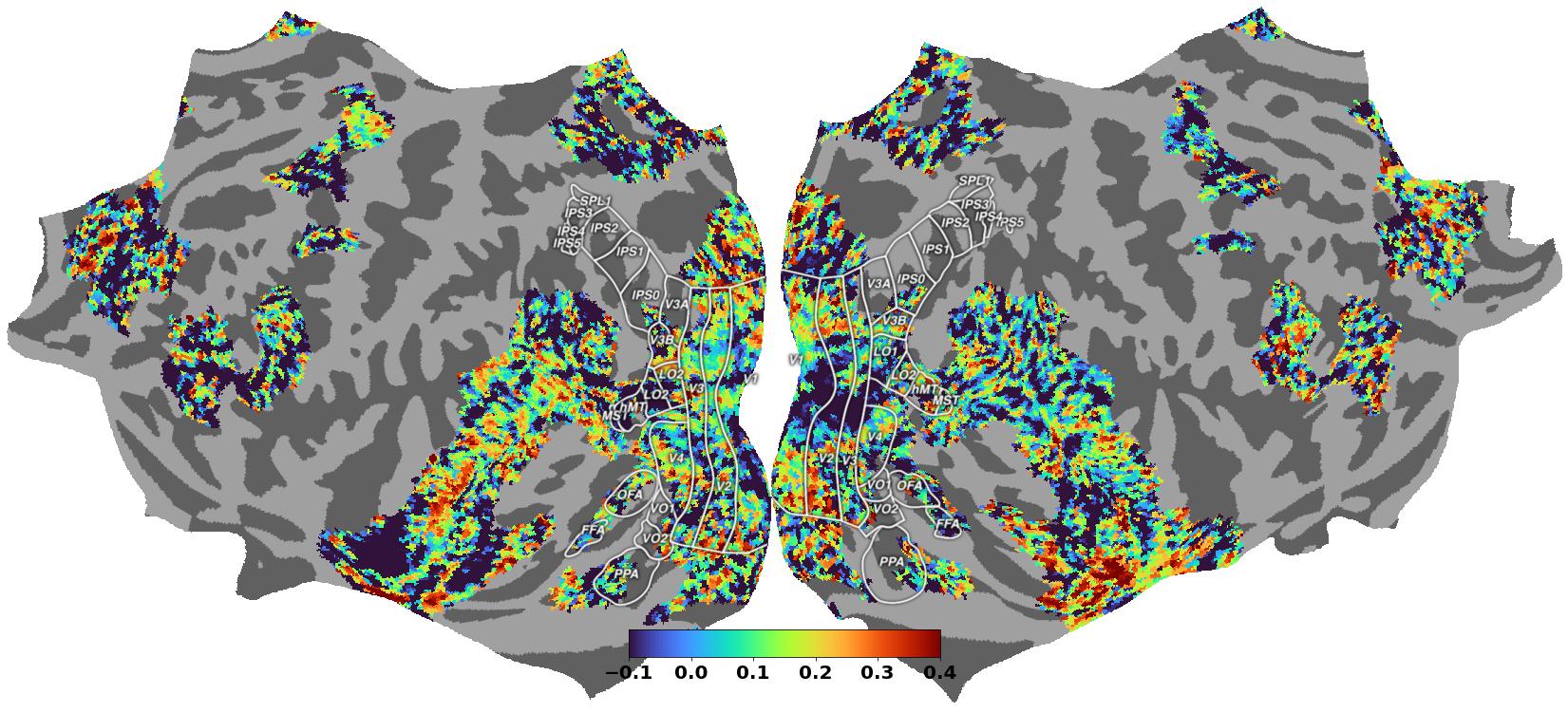}
    \\(c) Subject-05 \\
\end{minipage}
\caption{Estimated cross-subject prediction accuracy for all four participants for the naturalistic movie watching. Pearson correlation scores for each voxel in each subject are projected onto the subject’s flattened cortical surface.}
\label{fig:noise_ceiling_subjects}
\end{figure*}

Fig.~\ref{fig:noise_ceiling_subjects} shows the estimated cross-subject prediction accuracy for all four participants for the naturalistic movie watching. Pearson correlation scores for each voxel in each subject are projected onto the subject’s flattened cortical surface. The plots show that across all subjects higher activity is observed in the language and visual regions with a max correlation up to 0.4 implying that data has low noise and low cross-subject variability.

\section{Description of Narrative video understanding tasks}
\label{app:narrative_tasks_description}
Each task addresses a unique aspect of narrative meaning. Character Motivation captures implicit beliefs and goals of agents \cite{rabinowitz2018machine}. Event Boundary Detection identifies fine-grained change points critical for temporal segmentation \cite{zacks2007event}. Multi-Scene Summary integrates local events across broader spans. Narrative Summary consolidates long-range relational structure and thematic coherence \cite{bartlett1995remembering}. Removing any one task leaves certain forms of information unrecoverable from the others (e.g., agent beliefs, change points, relational links, or fine event details), establishing necessity. Conversely, under the decomposition and representability assumptions, the four tasks together span the relevant semantic space, up to a simple readout, establishing sufficiency.

\setlength{\tabcolsep}{1.5pt}
\begin{table}[!ht]
\begin{center}
\scriptsize
\caption{Instructions for multimodal narrative understanding tasks. 
}
\label{prompt_instructions}
%\vspace{-0.1cm}
\resizebox{\textwidth}{!}{ \begin{tabular}{|p{0.15\columnwidth}|p{0.85\columnwidth}|}
\hline
\bf Task&\bf Description\\ 
\hline
Character Motivation& Based on this given video segment, describe the character's likely motivation or emotional state at this moment. Justify your answer using visual or contextual cues. \\  \hline
Event Boundary Detection& You are shown a video segment from a larger movie. Based only on this segment, does this moment likely represent a narrative event boundary, such as a scene change? \\  \hline
Multi Scene Summary& You are shown a video segment from a larger movie. Based only on this clip, explain how this moment might connect to the broader narrative. \\ \hline
Narrative Summary& Summarize what is happening in this video segment. Clearly describe the observed events, characters, and any inferences about the context. 
 \\ \hline
\end{tabular}}
\end{center}
\end{table}

\setlength{\tabcolsep}{1.5pt}
\begin{table}[!ht]
\begin{center}
\scriptsize
\caption{Paraphrased and generic instructions for multimodal narrative understanding tasks. 
}
\label{prompt_instructions_rephrased}
%\vspace{-0.1cm}
\resizebox{\textwidth}{!}{ \begin{tabular}{|p{0.15\columnwidth}|p{0.85\columnwidth}|}
\hline
\bf Task&\bf Description\\ 
\hline
Character Motivation& Watch this video clip from a movie. What is the character feeling or trying to achieve in this moment? Support your interpretation using cues from the clip \\  \hline
Event Boundary Detection& Watch this video clip from a movie. Does this moment appear to mark a transition in the story, such as a change in scene, time, or topic? Explain your answer using cues from the clip. \\  \hline
Multi Scene Summary& Watch this video clip from a movie. How does this moment fit into the broader story? Describe the narrative connection you can infer from the clip \\ \hline
Narrative Summary& Watch this video clip from a movie. Describe what happens in this moment, including the key actions, characters, and situation shown in the clip. 
 \\ \hline
Generic Prompt & You are shown a video segment from a larger movie. Summarize what is happening in this segment, clearly describing the observed events, characters, and any inferences about the context. Describe the character's likely motivation or emotional state at this moment, justifying your answer using visual or contextual cues. Explain how this moment might connect to the broader narrative. Finally, indicate whether this moment likely represents a narrative event boundary, such as a scene change. \\
\hline
\end{tabular}}
\end{center}
\end{table}

\section{Implementation details for reproducibility}
\label{app:TrainingDetails}
%\noindent\textbf{Implementation details for reproducibility.}
All feature extraction experiments were conducted on a machine equipped with an NVIDIA A100 GPU with 80 GB of GPU RAM, partitioned into two devices of 40 GB each. The voxelwise encoding models were trained on NVIDIA GeForce RTX 3050 GPU with 4GB of GPU RAM. We used banded ridge-regression with the following parameters: MSE loss function; L2-decay ($\lambda$) varied from  10$^{-1}$ to 10$^{3}$; the best $\lambda$ was chosen by tuning on validation data that comprised a randomly chosen 10\% subset from the train set used only for hyper-parameter tuning. 

\section{Statistical Significance}
\label{app:statSig}
To determine if normalized predictivity scores are  significantly higher than chance, we run a permutation test using blocks of 10 contiguous fMRI TRs (considering the slowness of hemodynamic response) rather than individual TRs. By permuting predictions 5000 times, we create an empirical distribution for chance performance, from which we estimate p-value of the actual performance. The choice of these specific permutation test configurations is based on established methodologies in previous research~\citep{deniz2019representation,reddy2021can,oota2024speech}. To estimate the statistical significance of performance differences, such as between the model's predictions and chance or residual predictions and chance, we utilized the Wilcoxon signed-rank test~\citep{conover1999practical}, applying it to the mean normalized predictivity for the participants. 
Finally, the Benjamini-Hochberg False Discovery Rate (FDR) correction for multiple comparisons~\citep{benjamini1995controlling} is used for all the tests (appropriate because fMRI data is considered to have positive
dependence~\citep{genovese2000bayesian}).

\FloatBarrier

\section{Window-wise Voxel Alignment for DATE and TimeSformer}
\label{app:window-wise-alignment-date-timesformer}
\begin{figure*}[!ht]
    \centering
    \includegraphics[width=0.7\linewidth]{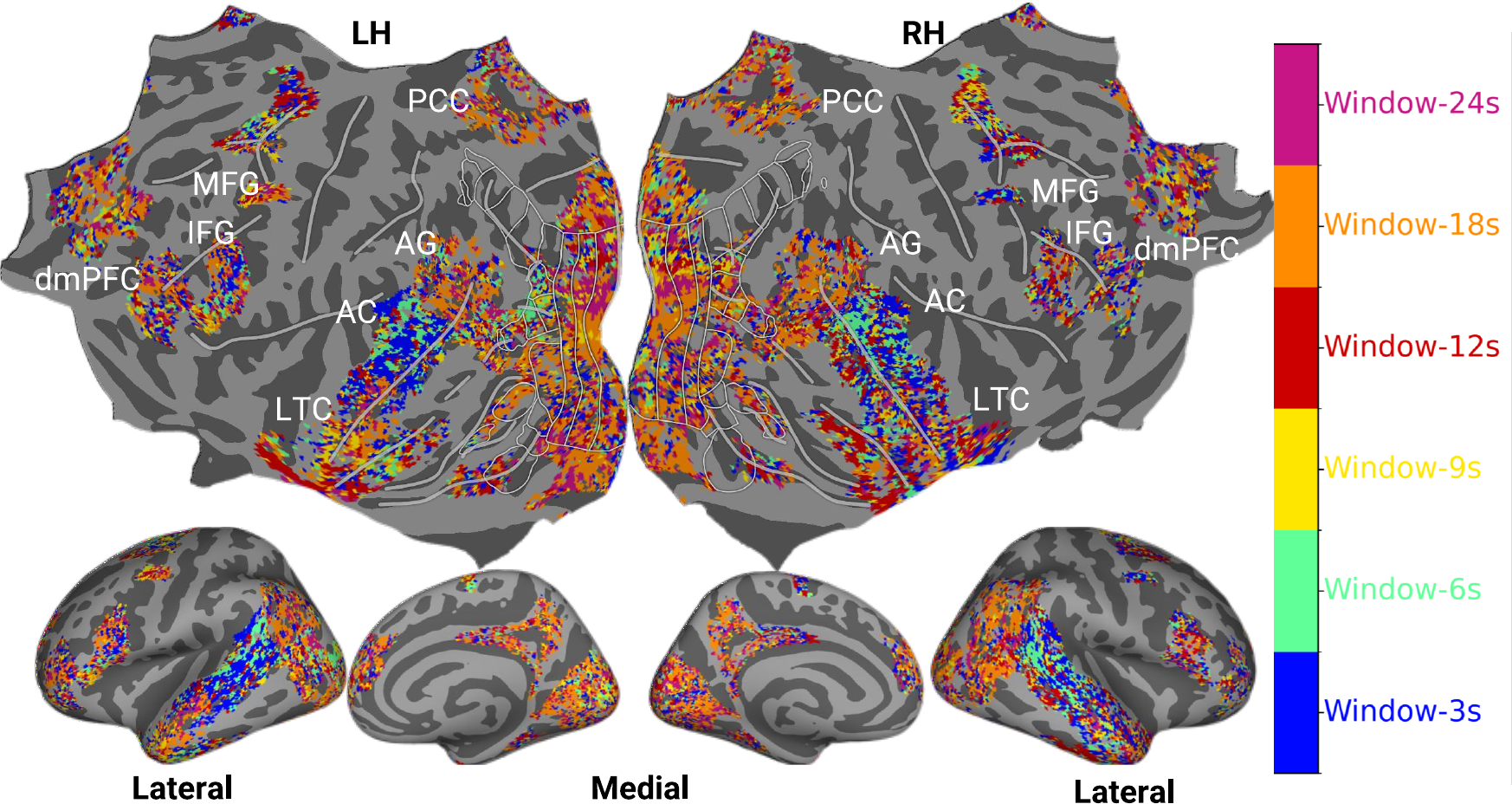}
    \includegraphics[width=0.7\linewidth]{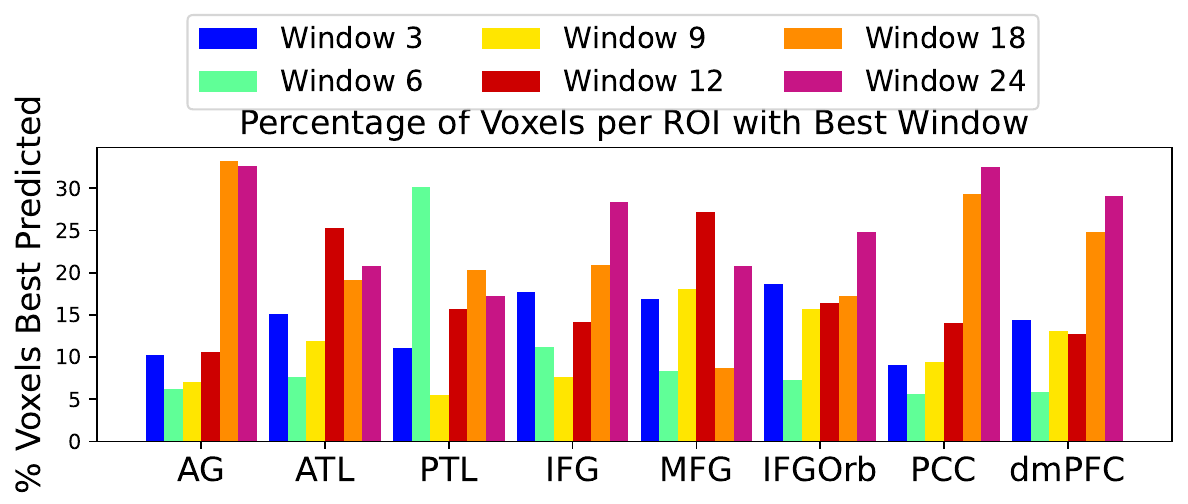}
    \caption{DATE model results: Each voxel is color-coded with the video duration (window length) that led to the highest normalized brain alignment with the DATE model. The color bar highlights color codes for each window length. (Top) The voxels are projected onto the flattened cortical surface of the `fsaverage' subject.  
    % , with applied hex color codes for the 10 task instructions, 
     (Bottom): Percentage of voxels in each ROI corresponding to each window length.}    \label{fig:temporal_context_mllm_date}
\end{figure*}

To test whether the temporal-context effects generalize beyond Qwen-2.5-Omni, we repeat the same window-wise voxel analysis using the DATE model. As in the main analysis, for each voxel we select the temporal window length (3\,s, 6\,s, 9\,s, 12\,s, 18\,s, or 24\,s) that yields the highest normalized brain alignment, and visualize the resulting best-window map on the \texttt{fsaverage} surface along with the ROI-wise distribution of best windows (Appendix Fig.~\ref{fig:temporal_context_mllm_date}).

Qualitatively, as shown in Fig.~\ref{fig:temporal_context_mllm_date} (top), DATE exhibits a similar temporal gradient to Qwen-2.5-Omni: shorter-to-intermediate windows (3--6\,s) account for a larger share of best-predicted voxels in more perceptual and ``local'' regions, while longer windows (18-24\,s) dominate in higher-order semantic/integrative regions. In particular, regions such as PCC and dmPFC show a preference for the longest window, consistent with longer-timescale narrative integration. Likewise, several language-selective ROIs (e.g., ATL, IFG, MFG, and IFGOrb) show a clear shift toward longer context windows, whereas PTL retain a comparatively stronger contribution from shorter windows, while AG shows both shorter and longer windows dominance, suggesting that parts of the temporal language system are optimally predicted at shorter timescales. Visual and auditory ROIs show weaker monotonic shifts overall, with early visual cortex appearing comparatively mixed across windows.

\begin{figure*}[!ht]
    \centering
    \includegraphics[width=0.49\linewidth]{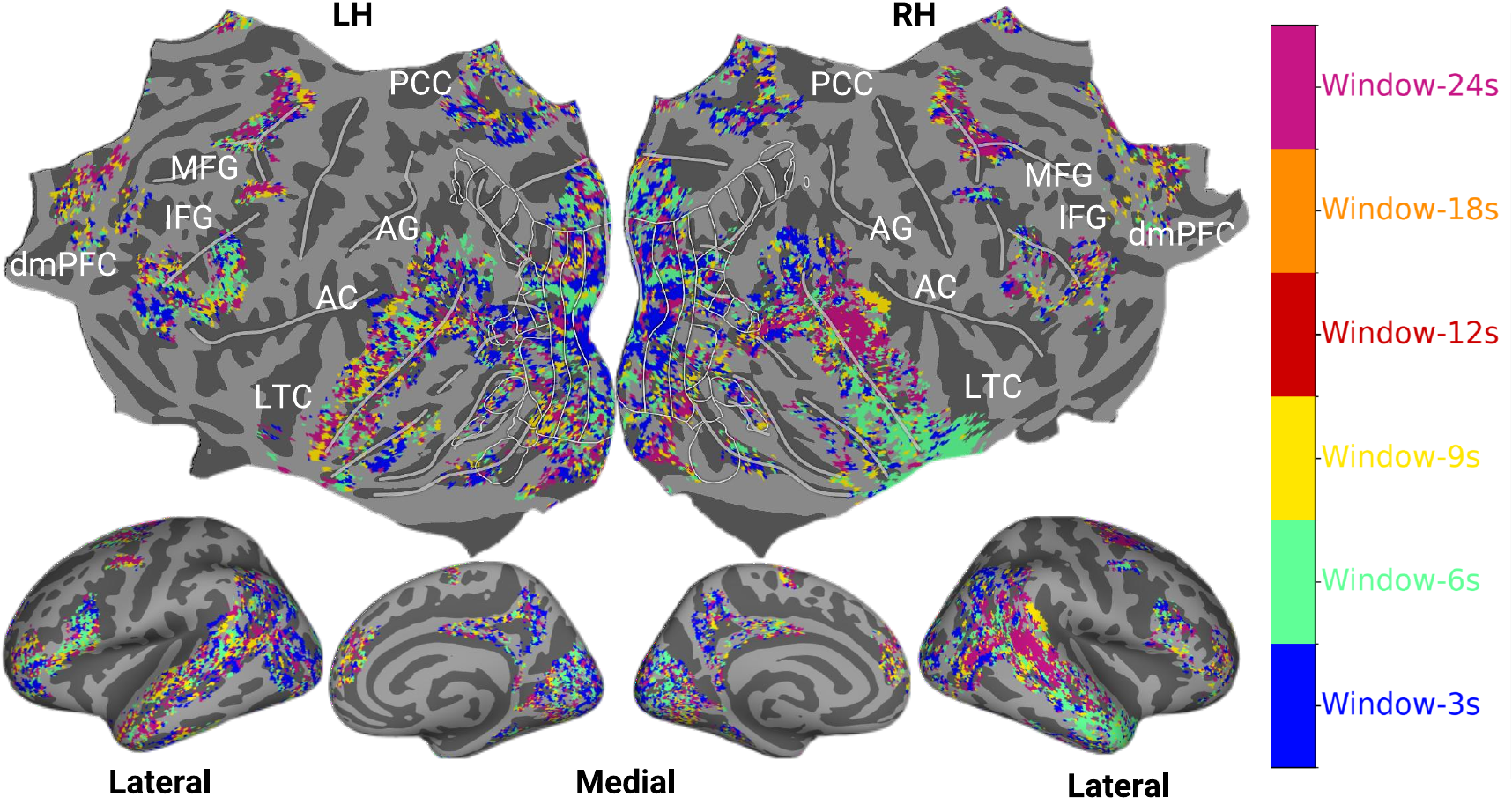}
    \includegraphics[width=0.49\linewidth]{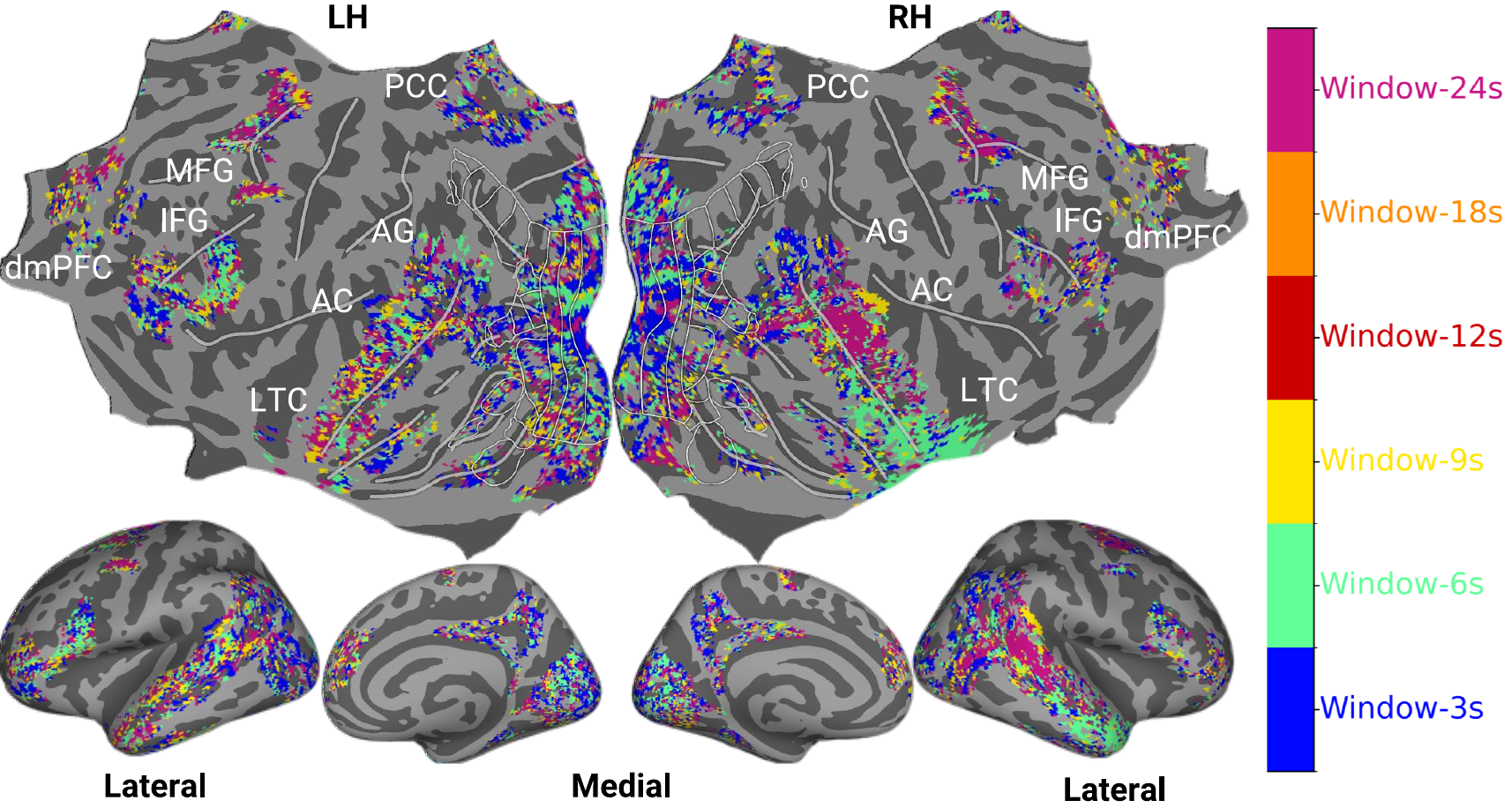}
    \caption{Unimodal video baselines (TimesFormer: Left, and VideoMAE: Right): Each voxel is color-coded with the video duration (window length) that led to the highest normalized brain alignment with the Unimodal video baselines. The color bar highlights color codes for each window length. (Left) The voxels are projected onto the flattened cortical surface of the `fsaverage' subject.}
    % , with applied hex color codes for the 10 task instructions, 
     %(Right): Percentage of best predicted voxels whose brain encoding performance is higher corresponding to each window length within language-selective regions, and visual regions on a flattened fsaverage cortical surface.}    
     \label{fig:temporal_context_timesformer}
\end{figure*}

\begin{figure*}[!ht]
    \centering
    \includegraphics[width=0.58\linewidth]{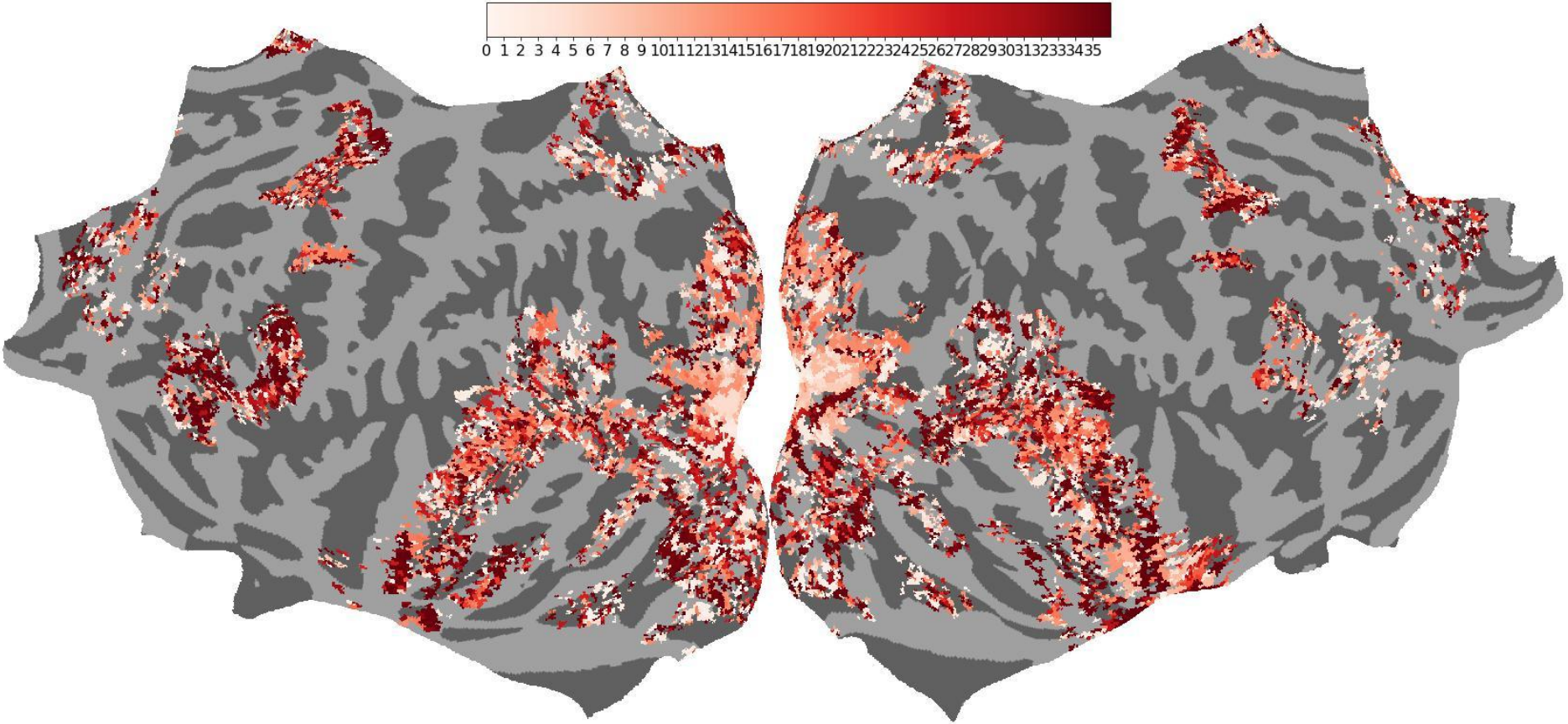}
    \includegraphics[width=0.36\linewidth]{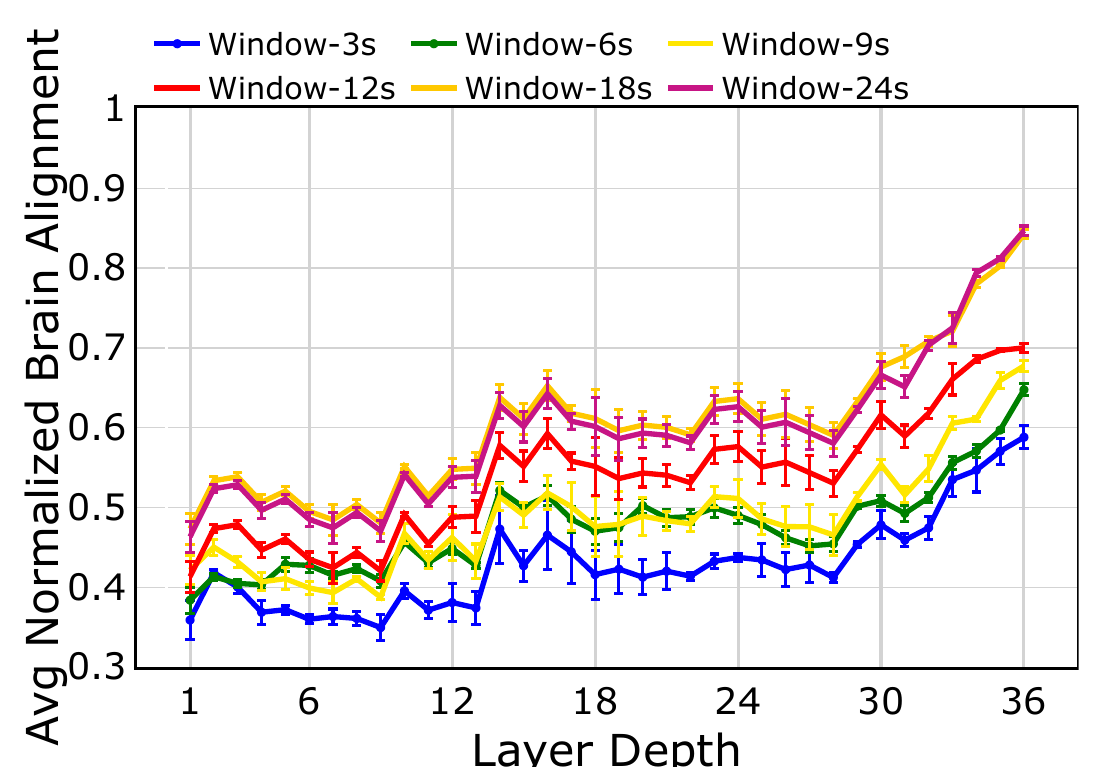}
    \caption{Layer-wise alignment results for DATE: Each voxel is color coded with the MLLM layer number (out of 36) that led to the highest normalized brain alignment. The color bar highlights color codes for each layer. 
    The voxels are projected onto the flattened cortical surface of the `fsaverage' subject.}
    \label{fig:qwen2.5-date_layers}
\end{figure*}

\begin{figure*}[!ht]
\centering
\begin{minipage}{0.6\textwidth}
\centering
    \includegraphics[width=\linewidth]{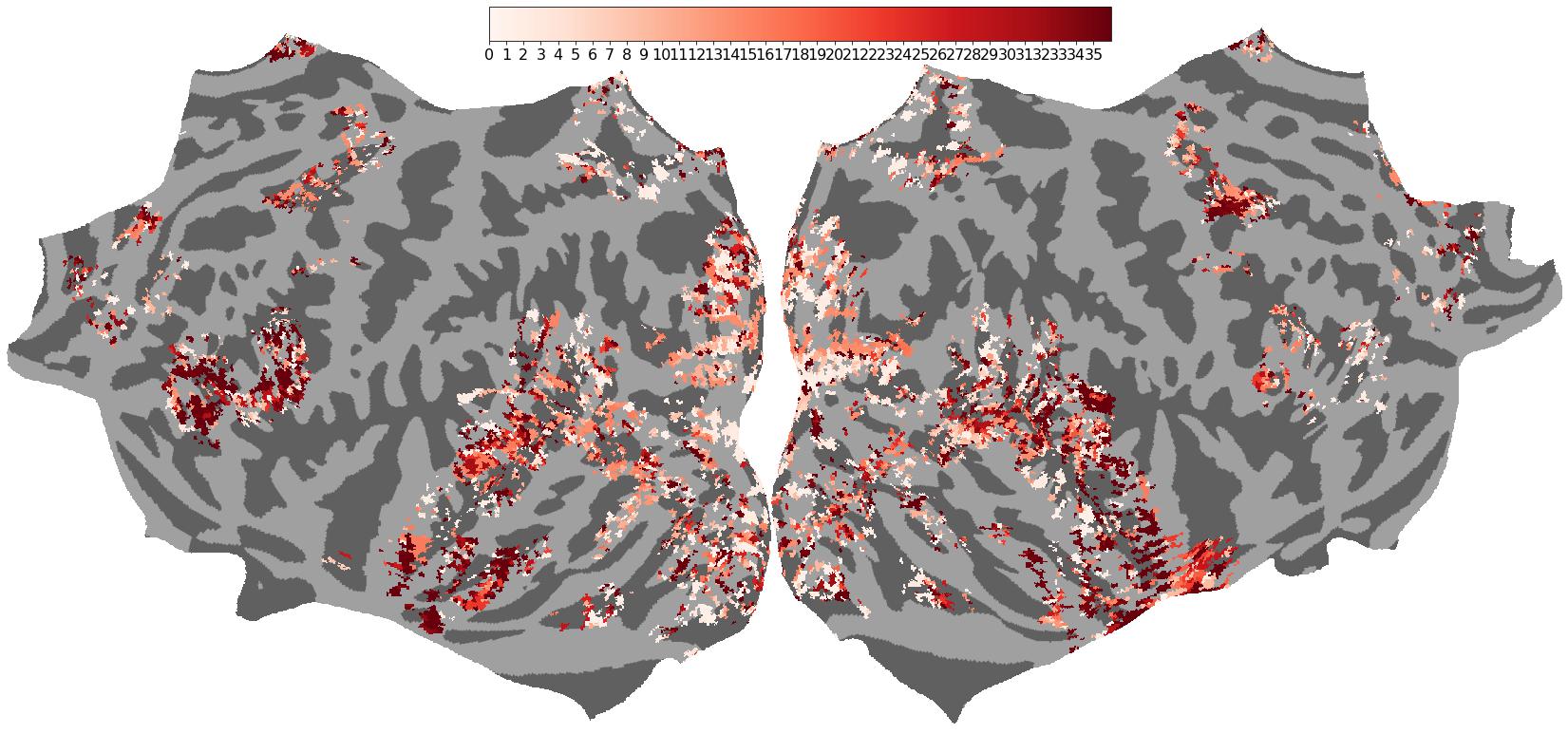}
    \\(a) Subject-01\\
\end{minipage}
\hfill
\begin{minipage}{0.6\textwidth}
\centering
    \includegraphics[width=\linewidth]{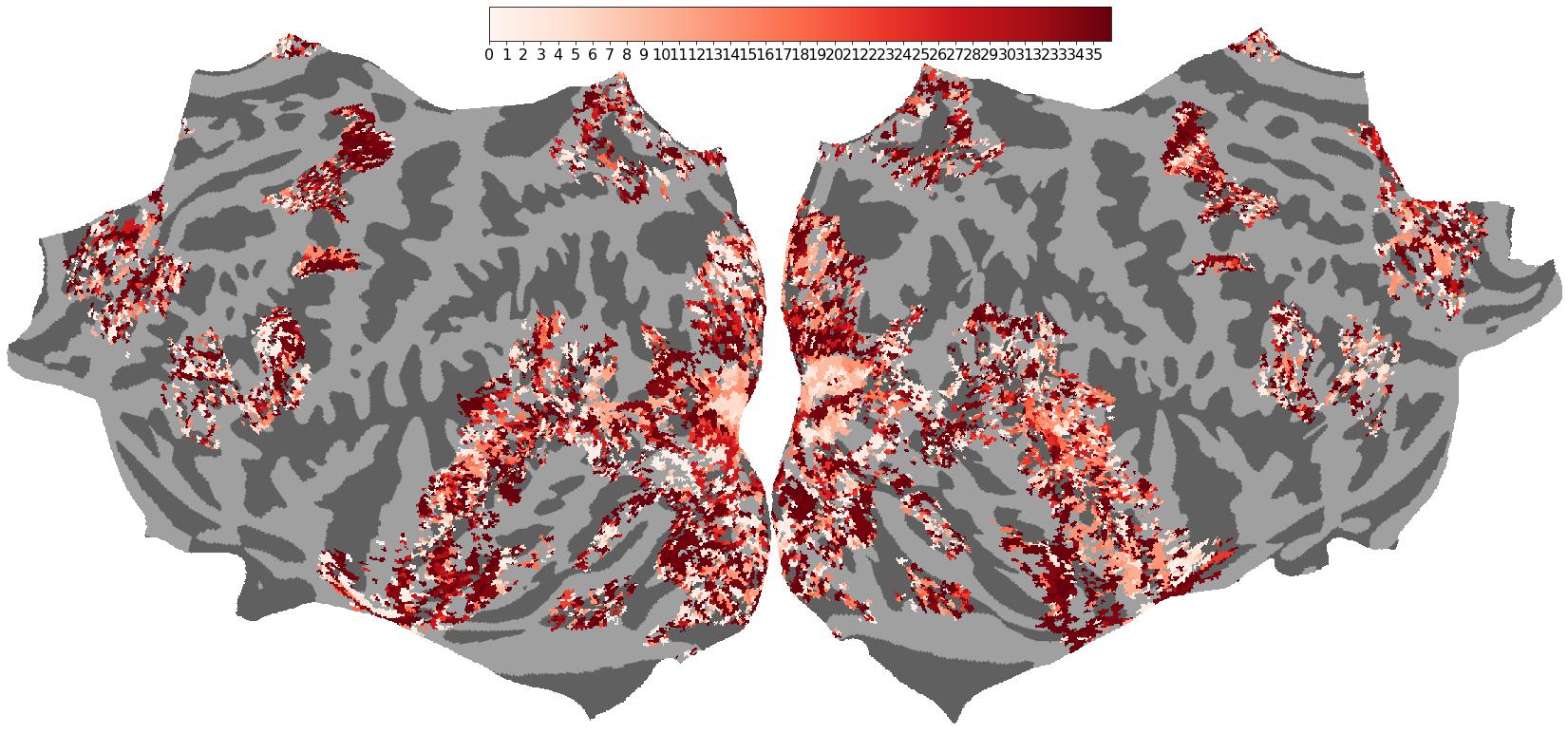}
    \\(b) Subject-02\\
\end{minipage}
\hfill
\begin{minipage}{0.6\textwidth}
\centering
    \includegraphics[width=\linewidth]{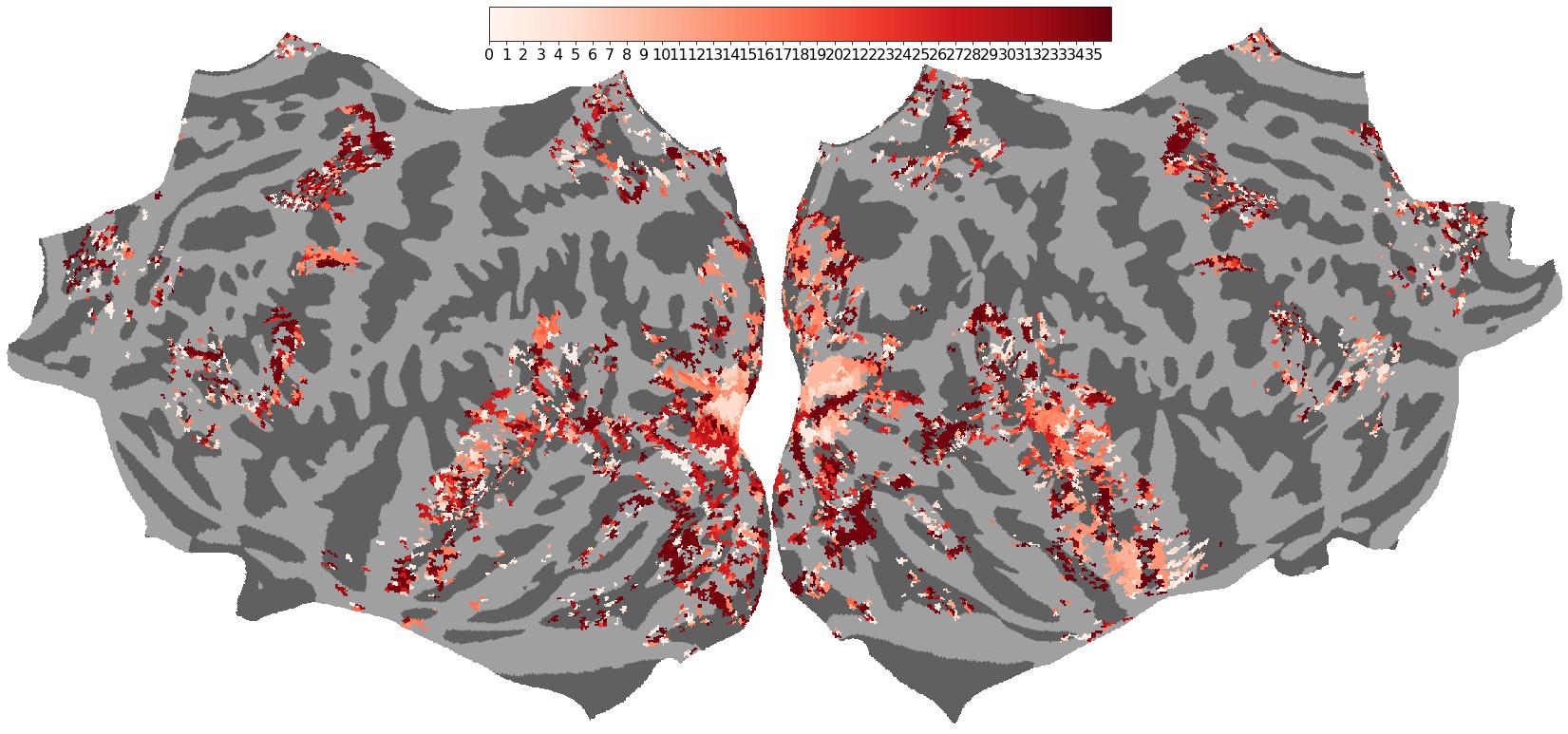}
    \\(c) Subject-03\\
\end{minipage}
\hfill
\begin{minipage}{0.6\textwidth}
\centering
    \includegraphics[width=\linewidth]{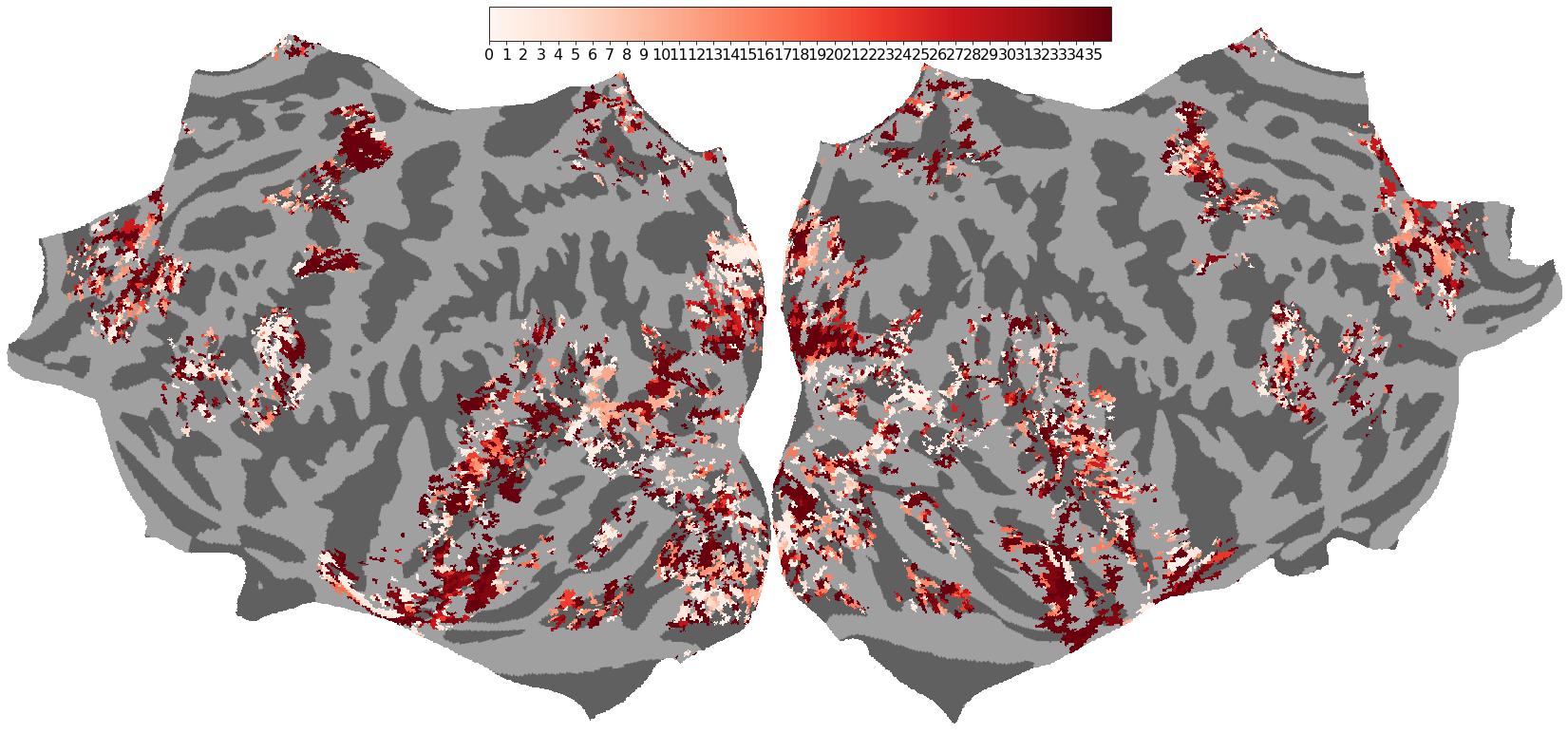}
    \\(d) Subject-05\\
\end{minipage}
    \caption{Qwen2.5-Omni: Each voxel is color coded with the video MLLM layer number (out of 36) that led to the highest
normalized brain alignment. The color bar highlights color codes for each layer. The voxels are projected onto the flattened cortical surface of each subject on `fsaverage' surface.}
    \label{fig:qwen_omni_layers_instruction_brainmap}
\end{figure*}
\begin{figure*}[t]
    \centering
    \includegraphics[width=0.7\linewidth]{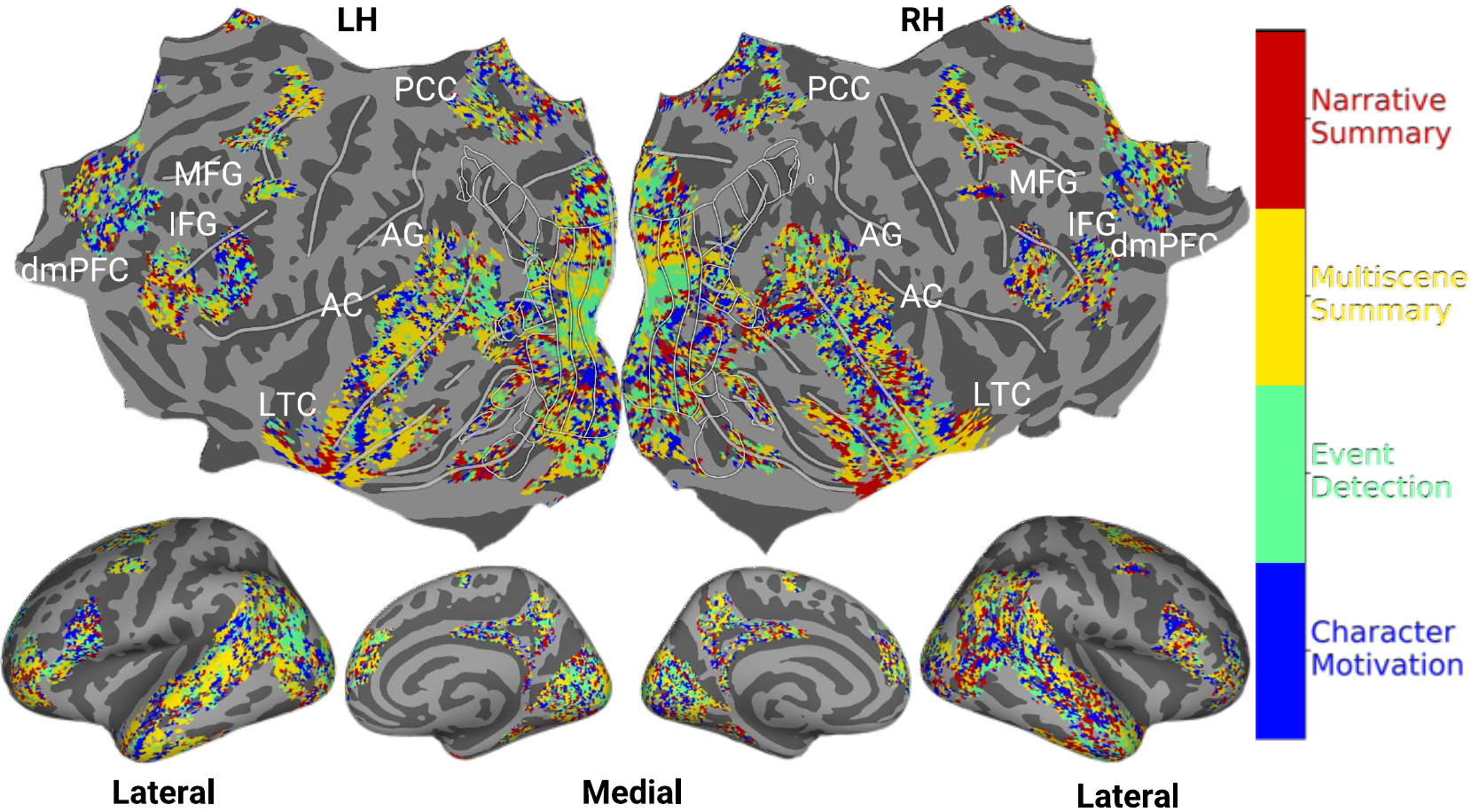}
    \includegraphics[width=0.6\linewidth]{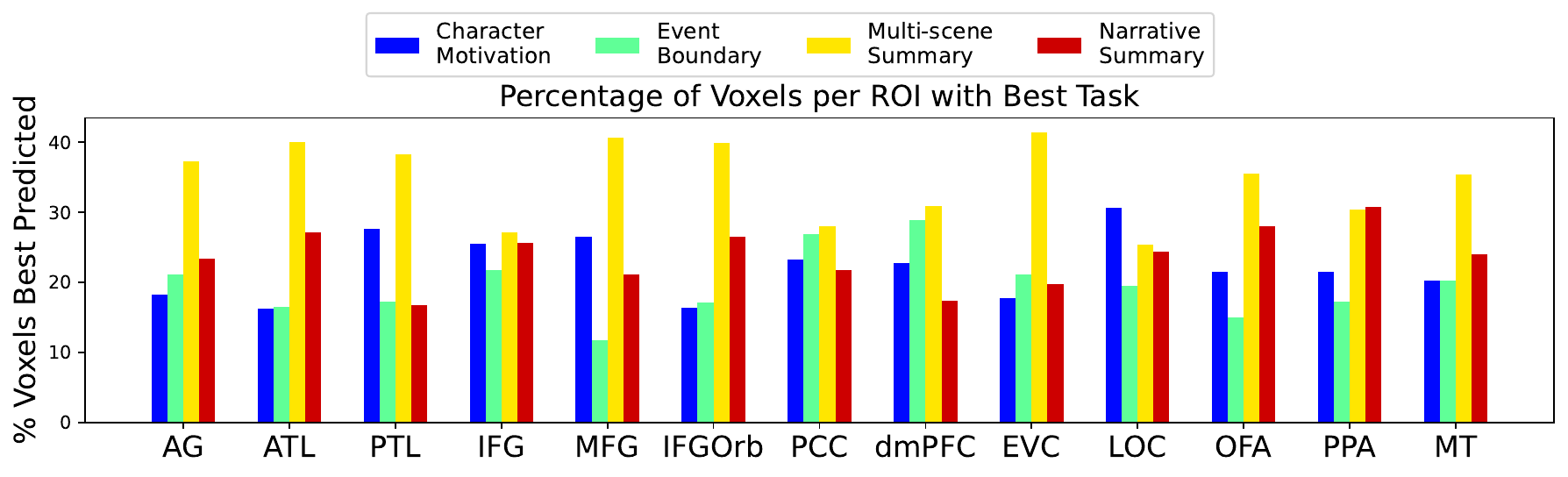}
    \caption{DATE model results: Each voxel is color-coded with the instruction that led to the highest normalized  with the DATE model. The color bar highlights color codes for each window length. (Top) The voxels are projected onto the flattened cortical surface of the `fsaverage' subject.  
    % , with applied hex color codes for the 10 task instructions, 
     (Bottom): Percentage of voxels in each ROI corresponding to each window length.}    \label{fig:tasks_mllm_date}
\end{figure*}
The ROI-wise histogram (Fig.~\ref{fig:temporal_context_mllm_date}, bottom) further supports these observations, showing that 18-24\,s is the most frequent best window across many higher-order ROIs, while 3--6\,s windows contribute more substantially in PTL/AG and several sensory-selective regions. Together, these results indicate that DATE captures heterogeneous temporal integration across cortex in a manner consistent with the main findings, reinforcing that long-context gains in brain alignment reflect model representations that better support long-timescale narrative processing rather than simply increased visual input length.

\section{Robustness to Frame Sampling.}
\label{app:frame_sampling}

Our default protocol uses a window-length-dependent frame count (9–30 frames; see Methodology in Sec.~\ref{sec:experimental_setup}), so effective frame density varies modestly across windows. To verify that the gains we report with longer context are not an artifact of this scheme, we re-ran the analysis under constant frame density (1,fps), in which the per-window frame count scales strictly with temporal span. Brain alignment increases monotonically with temporal span under both sampling strategies (Table~\ref{tab:fps_ablation}), and absolute differences between the two settings are small ($\sim$0.03–0.06). This indicates that the observed gains are driven primarily by temporal span rather than by frame count or density, and our main conclusion about the benefit of longer temporal context is not an artifact of the frame-sampling strategy.

\begin{table}[!ht]
\centering
\caption{Brain alignment under constant frame density (1\,fps) vs.\ our default scaled sampling (9, 9, 12, 16, 24, and 30 frames for 3, 6, 9, 12, 18, 24\,s). At 3\,s under the 1\,fps condition we use 4 frames, the minimum required by the MLLM input pipeline. The monotonic increase with temporal span is preserved under both strategies.}
\label{tab:fps_ablation}
\begin{tabular}{lcc}
\toprule
Window & 1\,fps$^{\dagger}$ & Default sampling \\
\midrule
3\,s  & 0.636 & 0.667 \\
6\,s  & 0.692 & 0.732 \\
9\,s  & 0.752 & 0.785 \\
12\,s & 0.818 & 0.843 \\
18\,s & 0.957 & 0.894 \\
24\,s & 0.943 & 0.890 \\
\bottomrule
\end{tabular}
\\[2pt]
{\footnotesize $^{\dagger}$3\,s uses 4 frames (model minimum); 6, 9, 12, 18, 24\,s use 6, 9, 12, 18, 24 frames.}
\end{table}

\section{Layer-wise Analysis for DATE}
\label{app:layer-wise-date-timesformer}

We investigate how MLLM brain-encoding performance varies across model layers as the temporal context window increases, as shown in Fig.~\ref{fig:qwen2.5-date_layers} (right). Across all temporal windows, we observe a consistent stratification into three layer groups: early layers (1--12), middle layers (13--28), and late layers (29--36). Overall, normalized brain alignment increases with longer sliding-window context, indicating that MLLMs benefit from long-range temporal information compared to shorter windows (e.g., 3\,s and 6\,s). We also find a layer-wise hierarchy: alignment generally improves from early to deeper layers across windows, suggesting progressively more abstract representations in later layers.

Fig.~\ref{fig:qwen2.5-date_layers} (left) visualizes voxel-wise layer preferences for the 12\,s condition, projected onto the \texttt{fsaverage} subject. We observe that (i) early sensory regions (early visual and auditory cortex) align best with lower layers, consistent with shallow representations capturing low-level sensory features; (ii) higher-level visual regions such as lateral occipital complex (LOC) and parahippocampal place area (PPA), and temporal/parietal language ROIs (PTL and AG) tend to align with middle-to-late layers; and (iii) language-related regions such as inferior frontal gyrus (IFG), anterior temporal lobe (ATL), and angular gyrus align most strongly with the deepest layers. Together, these results suggest that MLLMs exhibit a layered representational hierarchy that broadly mirrors cortical processing hierarchies.

\section{Task-wise Voxel Alignment for DATE}
\label{app:task-wise-alignment-date-timesformer}
To test whether the task-specific dissociations generalize beyond Qwen-2.5-Omni, we repeat the voxel-wise winner-task analysis using the DATE model. For each voxel, we select the narrative-task instruction that yields the highest normalized brain alignment and visualize the resulting winner-task maps on the \texttt{fsaverage} surface (Fig.~\ref{fig:tasks_mllm_date}) (top). We additionally quantify, for each ROI, the percentage of voxels for which each task is optimal (Fig.~\ref{fig:tasks_mllm_date} (bottom)).

Fig.~\ref{fig:tasks_mllm_date} (top) shows that DATE exhibits a qualitatively similar task-based organization: (i) Both \textit{Narrative Summary} and \textit{Multi-scene Summary} explain a large fraction of voxels across widespread association cortex, consistent with these prompts emphasizing long-range semantic integration; (ii) \textit{Character Motivation} more frequently emerges in lateral temporal and occipito-temporal regions, suggesting stronger alignment with regions supporting person- and situation-level inference grounded in audiovisual content; and (iii) \textit{Event Boundary} is comparatively sparse overall, but appears more prominently in select higher-order territories, indicating more specialized and localized boundary-related tuning rather than broad dominance.

The ROI-wise histogram (Fig.~\ref{fig:tasks_mllm_date} (bottom)) further supports these observations. \textit{Multi-scene Summary} is the plurality winner across many ROIs (including several language/association ROIs such as AG/ATL/MFG/IFGOrb, and multiple visual ROIs), indicating that multi-scene integration prompts capture representations that are broadly predictive across networks. In contrast, \textit{Narrative Summary} shows comparatively stronger contributions in some integrative ROIs (e.g., PPA and several frontal/temporal language ROIs), suggesting that global narrative compression preferentially matches higher-order semantic integration demands. \textit{Character Motivation} accounts for the largest share of best-predicted voxels in more ``local'' regions such as LOC (and a substantial fraction in PTL), consistent with character/agent-centric prompts aligning with regions sensitive to entity-level cues. Finally, \textit{Event Boundary} contributes relatively more in dmPFC and PCC than in most other ROIs, consistent with boundary-related structure being distributed but not the dominant explanatory factor in most regions.

Overall, DATE reproduces the main qualitative conclusion that narrative-task prompting acts as a functional probe: different prompts yield distinct, ROI-specific patterns of brain alignment rather than a single task dominating everywhere.

\section{Interpretability Analysis for Qwen2.5-Omni and DATE}
\label{app:Interpretability-Analysis-date-timesformer}

% \noindent\textbf{Similarity of voxel-wise encoding weights across tasks vs window lengths}

\begin{figure*}[t]
    \centering
    \includegraphics[width=\linewidth]{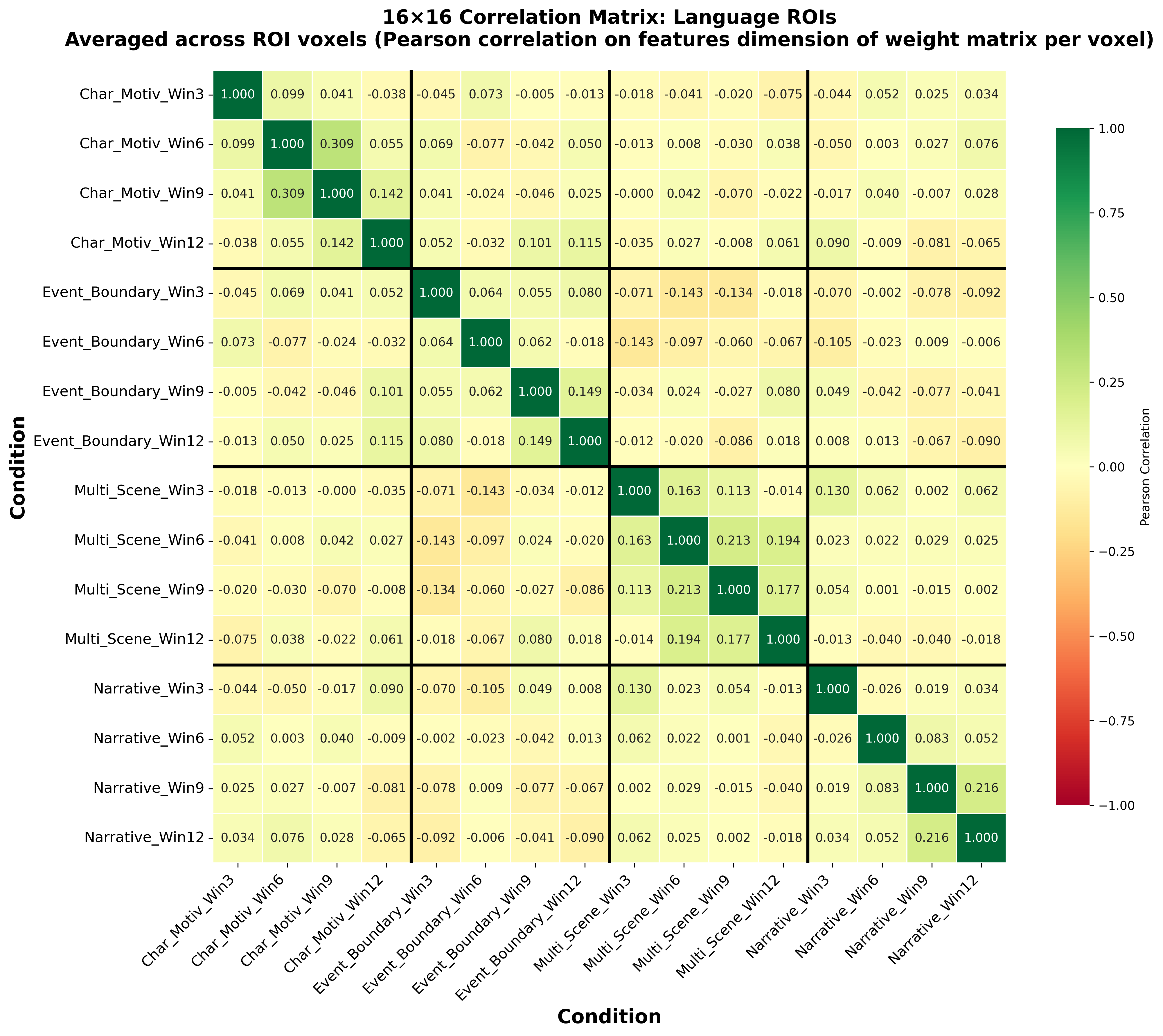}
    \caption{A 16×16 similarity matrix comparing the voxel-wise encoding weight patterns across 4 tasks × 4 temporal windows within language and visual voxels.}
    \label{fig:similarity_windows_tasks}
\end{figure*}

Fig.~\ref{fig:similarity_windows_tasks} shows a 16×16 similarity matrix comparing the voxel-wise encoding weight patterns across 4 tasks × 4 temporal window lengths within language and visual voxels. We make the following observations: (i) Task identity drives the largest differences in voxel-wise tuning (cross-task similarity is low). (ii) Increasing temporal context tends to make weight patterns more consistent within a task, especially for multi-scene integration and narrative, while event boundary remains relatively weakly consistent across windows. (iii) Even within a task, correlations are not high, implying that context length meaningfully reshapes the learned voxel-feature mapping rather than merely scaling it.

\section{Limitations}
\label{app:limitation}

Our study has several limitations that scope its claims and motivate future work. 
One possible limitation our study lies in the maximum temporal context: Prior brain-MLLM alignment work uses windows of only up to $\sim$1.49 s, whereas we substantially extend this to 3–24 s,  sufficient to establish the ROI-specific temporal gradient in Sec.~\ref{sec:results} and consistent with psycholinguistic MEG findings that mPFC and IFG integrate context for story-level meaning over longer timescales, a pattern we also observe in our long temporal-context experiments. Naturalistic narrative comprehension is argued to further engage windows of tens of seconds to a minute, which we leave to future work for compute reasons rather than scientific ones: feature extraction with video-audio MLLMs scales rapidly with sequence length and currently bottlenecks minute-scale brain-alignment analyses. Future work will focus on memory-efficient long-context MLLMs and hierarchical chunking are a clear next step that our findings directly motivate.
A second limitation is dataset scope: we use Movie10 from the Courtois NeuroMod project~\citep{st2023cneuromod}, which provides high-quality fMRI but is restricted to four publicly available subjects and English-language Hollywood-style films. Extending the analysis to other languages, narrative genres, and higher-temporal-resolution acquisition protocols (e.g., MEG/ECoG) is a natural follow-up.

\end{document}